%% file: main.tex
\documentclass[10pt,twocolumn,a4paper]{article}

\usepackage{graphicx}     
\usepackage{xspace}     
\usepackage{balance}     
\usepackage{booktabs}     
\usepackage{multi row}     
\usepackage[font={bf}, tableposition=top]{caption}     
\usepackage[hyphens]{url}     
\usepackage{bold-extra}     
\usepackage[vlined,linesnumbered,ruled,noend]{algorithm2e}     
\usepackage{microtype}    
\usepackage{units}     
\usepackage{mathtools}     
\usepackage{amssymb}     
\usepackage{subcaption}	
\usepackage{mathptmx} 

\graphicspath{{figures/}}

\newcommand{\spara}[1]{\smallskip\noindent\textbf{#1}}
\newcommand{\mpara}[1]{\medskip\noindent\textbf{#1}}

\newenvironment {squishlist}
{\begin{list}{$\bullet$}
  { \setlength{\itemsep}{0pt}
     \setlength{\parsep}{3pt}
     \setlength{\topsep}{3pt}
     \setlength{\partopsep}{0pt}
     \setlength{\leftmargin}{1.5em}
     \setlength{\labelwidth}{1em}
     \setlength{\labelsep}{0.5em} } }
{\end{list}}

\usepackage[utf8]{inputenc}
\usepackage{color}
\usepackage[show]{chato-notes}

\usepackage[greek,english]{babel}

\newcommand{\name}{{\scshape Aion}\xspace}
\newcommand{\namemeaning}{Greek god of eternity, personifying unbounded time.\xspace}
\newcommand{\caching}{\emph{proactive caching}\xspace}
\newcommand{\gc}{\emph{predictive cleanup}\xspace}
\newcommand{\pbucket}{m-bucket\xspace}
\newcommand{\pbuckets}{m-buckets\xspace}
\newcommand{\sbucket}{p-bucket\xspace}
\newcommand{\sbuckets}{p-buckets\xspace}

\begin{document}

\date{}

\title{Aion: Better Late than Never in Event-Time Streams\thanks{Aion ({\greektext Aiwn}) \namemeaning}
}

\author{
Sergio Esteves\\\small{INESC-ID}\\\small{Instituto Superior Técnico}\\\small{Universidade de Lisboa, Portugal}
\and 
Rodrigo Rodrigues\\\small{INESC-ID}\\\small{Instituto Superior Técnico}\\\small{Universidade de Lisboa, Portugal}
\and
Luís Veiga\\\small{INESC-ID}\\\small{Instituto Superior Técnico}\\\small{Universidade de Lisboa, Portugal}
\\\\\and
Gianmarco De Francisci Morales\\\small{Qatar Computing Research Institute}\\\small{Qatar}
\and
Marco Serafini\\\small{College of Information and Computer Sciences}\\\small{University of Massachusetts Amherst, USA}
}

\maketitle

\begin{abstract}
\input{sections/abstract}
\end{abstract}

\section{Introduction}
\label{sect:introduction}
\input{sections/introduction-2}

\section{Background}
\label{sect:background}
\input{sections/background}

\section{\name Design}
\label{sect:architecture}
\input{sections/architecture}

\section{Experimental Evaluation}
\label{sect:evaluation}
\input{sections/evaluation}

\section{Related Work}
\label{sect:related-work}
\input{sections/related-work}

\section{Conclusion}
\label{sect:conclusion}
\input{sections/conclusion}


\bibliographystyle{abbrv}
{\scriptsize \bibliography{refs}}

\end{document}

%% file: sections/abstract.tex
Processing data streams in near real-time is an increasingly important task. In the case of event-timestamped data, the stream processing system must promptly handle late events that arrive after the corresponding window has been processed. To enable this late processing, the window state must be maintained for a long period of time. However, current systems maintain  this state in memory, which either imposes a maximum period of tolerated lateness, or causes the system to degrade performance or even crash when the system memory runs out.


In this paper, we propose \emph{\name}, a comprehensive solution for handling late events in an efficient manner, implemented on top of Flink. In designing \name, we go beyond a naive solution that transfers state between memory and persistent storage on demand. In particular, we introduce a \caching scheme, where we leverage the semantics of stream processing to anticipate the need for bringing data to memory. Furthermore, we propose a \gc scheme to permanently discard window state based on the likelihood of receiving more late events, to prevent storage consumption from growing without bounds.

Our evaluation shows that \name is capable of maintaining sustainable levels of memory utilization while still preserving high throughput, low latency, and low staleness.



%% file: sections/introduction-2.tex

 Stream Processing Systems (SPS) are  increasingly employed to extract insights and value from continuous streams of data in near real-time. Examples of this class of systems include Storm~\cite{storm}, Spark Streaming~\cite{spark}, Samza~\cite{samza}, Apex~\cite{apex}, Google Cloud Dataflow~\cite{googleclouddataflow}, or Flink~\cite{DBLP:journals/debu/CarboneKEMHT15}. In many jobs handled by SPSs, each record in the stream represents a specific \emph{event}, which is associated with an \emph{event time}, e.g., a user clicked on an ad at a certain time. In these cases, events may arrive out of the order by which they were generated, and they are typically aggregated in event time windows (e.g., all clicks generated in the last hour), and subsequently processed by the SPS at a given \emph{processing time}.

The challenge with this model is that some events may experience large delays between generation and processing times. This can happen for a variety of reasons, such as network congestion, partitions, failures, configuration errors, or transient connections on the device generating the event. These delays can prevent events from arriving in time to be processed in their pertaining windows. 

The way that existing SPSs handle this case can be split into two categories. Some systems handle this by simply dropping late events (i.e., load shedding~\cite{Tu:2006:LSS:1182635.1164195}). However, dropping events is not acceptable in mission or business critical applications that rely on complete result sets (e.g., fraud detection, traffic monitoring, or intensive care units). For example, Google's Photon system~\cite{Ananthanarayanan:2013:PFS:2463676.2465272} is used for ad billing, and, as described, each time an ad click is permanently ignored due to delays, money is actually lost.  Hence, Google needs to set a very large threshold for ignoring late events (of the order of days), making such occurrence virtually impossible~\cite{Ananthanarayanan:2013:PFS:2463676.2465272}. Similarly, applications that log financial transactions may be forced to ensure that all events are incorporated in a given computation irrespective of their arrival time, in order to meet accounting and legal constraints.

Alternatively, other SPSs, such as Flink, allow late events to be aggregated in an expired window for an extended period of time. However, for applications with operators whose state increases monotonically with the ingested data (namely most user-defined functions), maintaining concurrent windows for considerably long periods of time can create a large memory pressure. In fact, it has been shown that SPSs are not equipped to deal with an unbounded space cost: they start thrashing with OS paging, perform excessive JVM garbage collection, or simply crash when they run out of memory~\cite{10.1007/978-3-642-10424-4_16}.

In this paper, we propose \name, a comprehensive solution to handle late arriving events in stream processing. \name is capable of managing window state across memory and persistent storage (e.g., HDD, SSD, NAS), while maintaining low latency and sustainable memory utilization. 

Designing \name required addressing several research challenges: how to manage state across disk to alleviate memory pressure while not introducing major penalties in the processing rate due to I/O; for how long should the state of a window be maintained by the SPS; and how to update and refine results in a timely and resource-efficient manner.
\name tackles these challenges by introducing several key techniques that leverage the semantics of stream processing in order to improve the management of data across memory and persistent storage. 

The first technique is \caching, which treats main memory as a cache for the window state, which is otherwise offloaded to persistent storage.
The main insight of \caching is that the semantics of SPSs allow the system to predict that processing is more likely to be necessary at specific times, for example when a time window expires.
This enables using a {\em proactive} approach, where I/O is regulated by a central scheduler, which tries to evict data ahead of time, thus minimizing the performance penalty of offloading to persistent storage in terms of both latency and throughput
 
The second technique introduced by \name is \gc. This uses a past history of the distribution of late event arrival times to predict the best time to purge the state of a windowed operator completely, based on its likelihood of receiving more events; i.e., the state can be purged when we do not expect to receive more events (or, alternatively, less than a given fraction of  events for that window) within a chosen confidence interval.

Finally, we also address the issue of updating late results. For past windows, it is desirable to amend previously emitted results as soon as late events arrive. However, recomputing a monotonic window (whose state increases with ingested data) for each received event is computationally expensive. To address this, we provide a trigger that is able to find a good compromise between staleness of the result and resource usage (or number of executions), thereby identifying the adequate times for recomputing a past window.

We implemented \name by extending the codebase of Apache Flink, a widely used distributed SPS.
We  evaluate \name using benchmarks and practical applications.
Experimental results indicate that \name is capable of handling large amounts of lateness, well beyond the limit where current SPSs run out of memory and crash, thereby maintaining sustainable levels of memory utilization while still preserving high throughput, low latency, and low staleness.

The remainder of the paper is organized as follows. \S\ref{sect:background} provides the background and assumptions on stream processing necessary for later sections. \S\ref{sect:architecture} presents the design of \name. \S\ref{sec:impl} gives the main implementation details. \S\ref{sect:evaluation} presents our experimental evaluation. We survey related work in \S\ref{sect:related-work} and conclude in \S\ref{sect:conclusion}.

%% file: sections/background.tex

Before delving into the technical details of our system, we review a few key concept related to the computational semantics of SPSs supporting event-time processing.

Streaming applications are commonly represented in the form of directed graphs that represent the \emph{data flow} of the application.
The vertices of the graph are data transformations (operators), and its edges are channels that route data between operators.
The data flowing along these edges is a stream, represented as a sequence of \emph{events}, each associated with a \emph{key} and a \emph{timestamp}.
The key is specified by the application, and is data-dependent (e.g., an ad identifier).
To achieve high throughput, modern distributed engines leverage data parallelism by creating several instances of an operator that process independent sub-streams.

An SPS reads data from one or more sources. The rate at which the data is read is called \emph{ingestion rate}, 
whereas the rate at which an operator processes data is called \emph{processing rate}.
For an SPS deployment to be sustainable, it needs to offer a processing rate that can cope with the ingestion rate, at least on average over time.

\spara{Time domains.}
An important component of the abstraction provided by the operators is that events are associated with a timestamp. 
For assigning these timestamps, three different notions of time have been considered: \emph{processing-time}, \emph{ingestion-time}, and \emph{event-time}~\cite{Srivastava:2004:FTM:1055558.1055596}.

With processing-time, each operator assign a timestamp to an event independently, based on the the current system clock time when it processes the event.
Ingestion-time refers to the time when events enter the system, and is assigned to an event by the first operator that reads it from the data source.
%
Finally, event-time is associated to an event outside the SPS, when it is generated.
Event-time enables out-of-order streams of data events to be grouped and ordered by their timestamps, hence giving consistent results that are robust to delays (i.e., the result of the computation is the same irrespective of the order in which events are processed) or the mode of operation of the system~\cite{dataflow}. 


\spara{Windowing.}
A window groups events in time, allowing an infinite stream to be processed in finite batches~\cite{li2005semantics}.
A single event can be part of zero, one, or many windows, according to the user-specified semantics.
Common types of windows include: \textit{Tumbling}, fixed-size window with no overlap with other windows (e.g., to compute hourly aggregates); \textit{Sliding}, fixed-size window that slides by some amount (e.g., compute hourly aggregates every 10 minutes); \textit{Session}, dynamically sized window, which represents a consecutive, data-dependent portion of the stream, usually defined per key (e.g., a group of events separated in time by no more than a defined gap constant); and \textit{Count}, window that groups a fixed number of consecutive events, irrespective of their timestamps.


\spara{Watermarks.}
When using event-time, a \emph{watermark} signals the time when the system {\em assumes} that all events up to a certain event timestamp $t$ have arrived at an operator~\cite{jefferson1985virtual}.
For example, a watermark can signal that (ideally) all events in a given window have been received.
A watermark is always a best guess: events with a timestamp lower than the watermark timestamp $t$ may still be received, and are considered \emph{late}.
Late data may simply be dropped, or, in case the SPS can handle lateness, incorporated into the state of a window that has already been processed.
In the latter case, several semantics are possible, depending on the requirements of the application.
For instance, given an operator that computes the average of the values in a window, a late event might trigger the window to emit the new average incrementally by simply keeping track of the sum and the number of items, and updating these upon receiving late events. However, the case of non-linear functions such as percentiles or arbitrary UDFs is particularly complex: the whole state of the old window needs to be maintained in order to allow late events, and the whole computation needs to be re-executed. One of the design goals of \name is to be generic, thus handling such operators.

There are two main types of watermarks: periodic and punctuated.
Periodic watermarks are emitted based on either processing time (every $p$ seconds) or stream elements (every $p$ events).
In turn, punctuated watermarks are emitted based on conditions inferred from the data when a particular event arrives.
For instance, a source might emit a watermark when an explicit flush event arrives.

The generation of a watermark involves a delicate trade-off.
If it is emitted in a conservative way, the system might wait longer than actually needed to process events, thus increasing latency.
Conversely, if watermark is emitted too fast, a large fractions of events will become late, thus adding overhead to the computation.
Issuing watermarks is often based on a heuristic, since it is in general impossible to tell when all events belonging to a window have arrived~\cite{murray2013naiad}.




\spara{Triggering.}
A trigger is a mechanism that determines when a windowed operation should be executed, i.e., when to compute the value of the function over the data in the window.
By default, a window is triggered when its watermark is emitted, but it can also be triggered at other times, using different policies, e.g., percentile based (when some percentage of the data has been accumulated), data based (counts, punctuation, pattern matching), or even via external signals. 
In addition, for late events, a window is also triggered when the system time has reached the watermark plus the \emph{maximum allowed lateness}, which means that no further late events for the window are accepted, and the function result is final.

\spara{Operator State.}
The discussion so far can be applied to any modern SPS.
However, in terms of the operator semantics and the state they are able to maintain, there is no generally accepted API.
Therefore, we focus on the system that we use as a base for our implementation, which is Apache Flink.

State is used in stateful operators, which need to retain some memory of the events that were previously processed (e.g., counts for aggregates, parts of the streams for pattern matching, model parameters for machine learning).
Flink uses a managed state API, by which operators can access a set of standard state prototypes, usually one per key:

\begin{squishlist}
\item{\textbf{ValueState}}, a single value that can be retrieved and updated, e.g., a boolean indicating if an event with the same key has been received in the current window;
\item{\textbf{ReducingState/FoldingState}}, a single value that represents an aggregate of the processed sub-stream, computed via a reduce or fold function, e.g., a per-key sum of the events in the window;
\item{\textbf{ListState}}, a list of elements, which can be iterated and appended to, usually containing the events in the window;
\end{squishlist}

Of the three state prototypes, ListState is the one used by default in custom operators, as it is the most general.
However, it is also the most expensive in terms of memory, which can cause heavy pressure when the system needs to maintain a large number of windows active (because of a conservative watermark, or a large maximum allowed lateness).

Each operator can declare several state elements, and Flink will manage their distribution and lifecycle (checkpointing and restoring).
For each window processed by an operator, Flink maintains a separate instance of the operator state.

The default state backend of Flink stores the state in memory.
When the maximum allowed lateness for an operator is large, the number of windows to maintain can grow considerably, thus exerting pressure on the main memory.
When designing \name, our goal is to make judicious use of persistent storage to limit main memory usage, and thus alleviate this pressure, without sacrificing throughput or latency.

%% file: sections/architecture.tex

\name provides mechanisms to handle late events, by managing state data across both memory and persistent storage. In particular, our  goal  is to achieve the best of both worlds by ($i$) preserving the performance benefits of in-memory processing, while ($ii$) providing significantly more space to maintain state data across both main memory and persistent storage.

The way  \name is able to circumvent both memory size limitations and persistent storage latency is by taking advantage of the semantics of event-time stream processing, in order to perform a proactive management of past windows. In the remainder of this section, we outline how \name achieves these goals.


\subsection{Bucket Management}

\begin{figure}
	\includegraphics[width=\columnwidth]{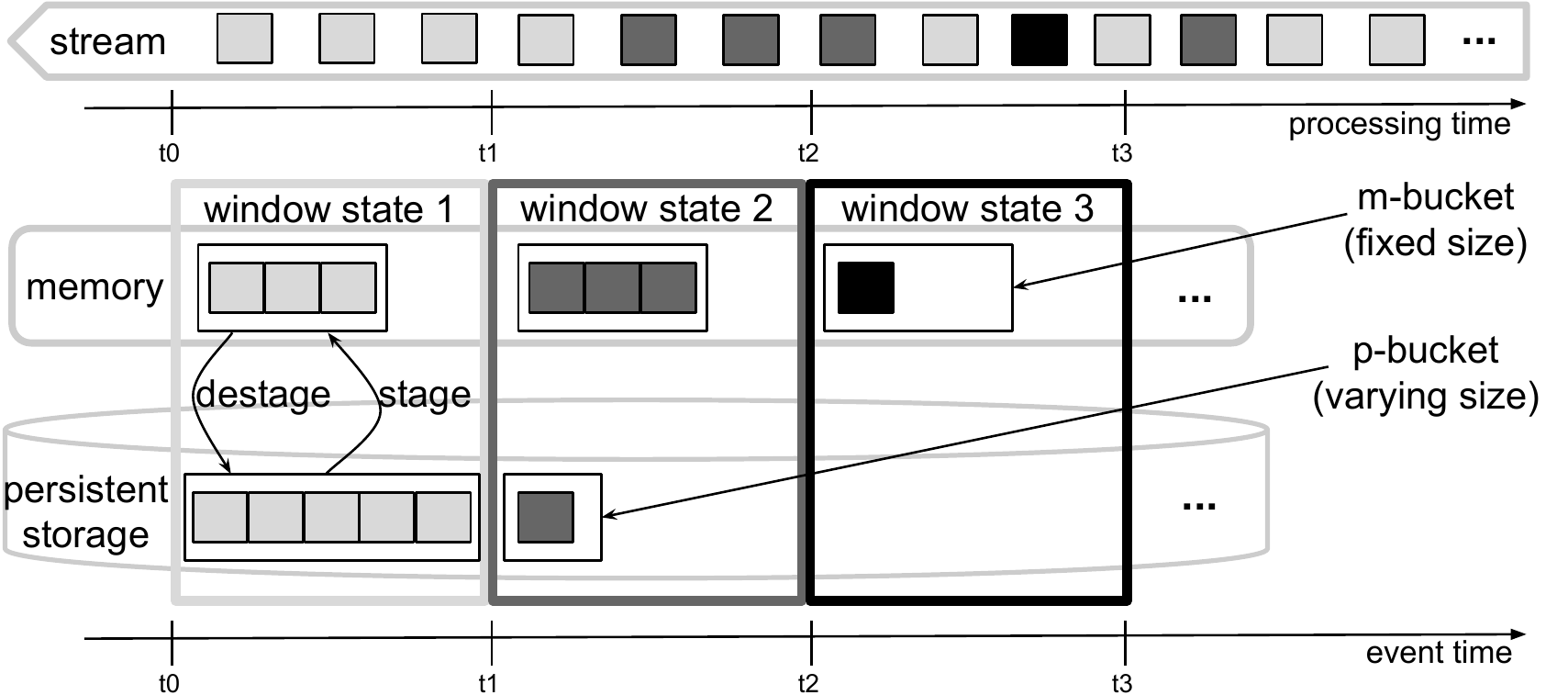}
    \caption{Different shades of gray represent how events are aggregated into windows. \name maintains the window state in both memory (\pbucket) and persistent storage (\sbucket).}
    \label{fig:architecture}
\end{figure}

\name splits the state of each window into two logical containers, called \emph{memory bucket} and \emph{persistent bucket} (abbreviated as \emph{\pbucket} and \emph{\sbucket}, respectively), as shown in Figure~\ref{fig:architecture}. The \pbucket resides in memory and has a limited maximum size; the \sbucket is in persistent storage and is only bounded by the total persistent storage size, which may be considerably larger (e.g., Terabytes).
\name keeps latency low by using \caching, which populates the \pbucket with events from the \sbucket, such that, in most cases, accessing in-memory data can be done without blocking on I/O.
%
In particular, this technique consists of transferring data between \pbuckets and \sbuckets in a way that is decoupled from the process of feeding window operators.
In other words, window operators always access \pbuckets, and the transfer of data from the \pbucket to the \sbucket and back is asynchronous.
This asynchrony then allows us to define flexible strategies for scheduling I/O, according to one of the policies that we explain next.

\subsection{Data Transfer Policies}

The choice of the timing of data transfers between the \pbucket and the \sbucket takes into consideration the semantics of the streaming application. 
In particular, there are four situations that \name needs to consider:
($i$) populating the window state for the first time;
($ii$) executing operators upon triggering;
($iii$) dealing with watermarks and integrating late events into past windows;
($iv$) computing final results.
We start by describing the standard policy for each of these situations, and then discuss several alternative policies.

\spara{Standard policy.}
When populating the state of window $w$, events are initially stored in the \pbucket of $w$.
When the \pbucket becomes full, \name redirects new events directly to the \sbucket. 
Subsequently, when $w$ is triggered for execution, \name executes the window operator by fetching all data from the associated \pbucket. At the same time, \name transfers data from the \sbucket to the \pbucket in the background, a process called {\em staging}.
Reading from the \pbucket while staging from the \sbucket allows us to mask the  I/O latency.

Eventually, the watermark reaches the end of $w$, which makes it expire (i.e., it becomes a past window). At this point, a \emph{destaging} operation takes place so that all data in the \pbucket is transferred to the \sbucket, thus releasing a significant amount of memory.
Subsequent arriving (late) events for \emph{w} are written directly to the \sbucket.

When a late event arrives, the window is scheduled for re-execution. However, the re-execution of a late window has low priority to avoid interfering with the execution of current windows, since these are the most up-to-date results that should be immediately displayed to the user. 

The key to reducing the I/O overhead associated with staging is to prestage state to the \pbucket before the re-execution occurs. To this end, we employ \caching, which estimates an appropriate time to start prestaging, by anticipating when the operator will re-execute. This is achieved by taking into account the different semantics of different types of watermarks. In particular, assessing re-execution time with periodic watermarks is trivial, since we have knowledge about the period of watermark generation and current logical time. In this case, during the first late execution for the window $w$, pre-staging starts pessimistically when the window immediately preceding $w$ fully expires (including maximum allowed lateness). During this process we assess the overall time taken ($\Delta t$) weighted by the number of staged events. Then, for subsequent re-executions of $w$, we start pre-staging $\Delta t$ time before the operator re-execution time.
For the case of punctuated watermarks, pre-staging for a window can start as soon as a late event for that window is received, since it indicates an upcoming re-execution, which may be delayed until pre-staging concludes. In both cases, the \pbucket of the past window is freed after re-execution.

\spara{Additional Policies.}
To be more flexible and extensible, \name's design allows for defining additional policies. They can be categorized as either \emph{local}, when they do not take into account the overall system memory utilization, or \emph{global}, in case they regard the system as a whole when optimizing memory. 

To demonstrate the flexibility of our design, we provide a few illustrative examples, starting with local policies:
\begin{itemize}
  \item When a watermark arrives, and if late events are allowed, destage the window state except for a (small) fraction $\rho_{min}$ of initial events, which act as a bootstrap set for later re-staging the window.
  \item When more than $\tau$ processing-time elapses (e.g., a multiple of median window processing time) without the window either getting new events or a watermark, destage the window state except the $\rho_{min}$ set.
\end{itemize}

Global policies, in turn, can include the following.
\begin{itemize}
\item When the available memory $\mu$ is moderately scarce, successively destage window state to disk (except their $\rho_{min}$ bootstrap set) in a selective way, e.g., either  by descending order of individual window state size (for faster savings), or by increasing values of ingestion rate of individual windows (to minimize  window processing delay) 
\item  When the available memory $\mu$ is very scarce (e.g., below a given threshold of 10\% of physical memory),  destage the state of all windows to disk except their $\rho_{min}$ set. 
\end{itemize}

\subsection{Sizing of the \pbuckets}
The size of the \pbuckets depends on the type of computation to be executed.
\emph{Blocking} window operators need to consume the entire input before starting the main processing task.
For example, when applying an FFT over a sliding window, the entire input data needs to be fetched before processing can start.
\emph{Non-blocking} window operators are able to perform the main processing task while events are fetched one by one.
This is the case, for example, when computing n-grams over a stream of words sorted by event time. 

For blocking operators, the size of the \pbucket should be equal to the size of the entire window input, otherwise computation is affected by I/O latency.
For non-blocking computations, the size of the \pbucket only matters, in terms of overall computation performance, when staging the events from the \sbucket takes longer than processing the events initially present in the \pbucket (thus, no longer masking I/O latency).
This constraint is driven by the relative sizes of the buckets and the relative speeds of staging and processing the events for a given computation. In summary, we aim for an \pbucket size that is large enough so that the function never has to wait for events that are still in the \sbucket.

\subsection{Permanently deleting window state}
\label{sect:trigger}

Ideally, \name should be able to handle unbounded lateness. However, not only  persistent storage is limited, but also the usefulness of windowed data becomes residual over large periods of time. Therefore, \name incorporates a \gc mechanism for purging  window state completely from the system, when that state is considered very unlikely to be needed.

The amount of elapsed time to perform \gc is updated in an adaptive way. To this end, the system continuously observes the distribution of late events (including late events that arrive beyond the maximum lateness bound). The idea is then to start with a conservatively large lateness bound, and, after a representative history of observations is collected, adjust this bound at runtime for newly created windows in a way that is estimated to cover a specified  percentage of the events (e.g., 99\%) within a certain confidence interval. We keep updating this distribution according to new observations, so that this estimate is as accurate and up-to-date as possible.

Before this maximum bound of allowed lateness expires, it is desirable to update previously emitted results once they become significantly inaccurate due to the arrival of new events. However, computing a window for each newly arrived late event can be very costly in terms of system resource usage. One possible solution to this problem would be to update this computation periodically; however, this can lead to unnecessary executions when the number of new events since the last execution is small or nonexistent. Conversely, during a spike of late events, the computed value might be significantly out-of-date for non-negligible periods. 


To address this, we introduce a new trigger that operates according to  \emph{staleness}. We define staleness, between pairs of consecutive executions, as $st = t * n / (T * N)$, where $t$ and $n$ are the time elapsed and the number of events accumulated since the last execution, respectively; $T$ and $N$ are the maximum possible time (i.e., maximum allowed lateness) and accumulated events (i.e., total number of late events expected), respectively. Staleness can be user defined, according to a specified SLA, e.g., a bound on the maximum outdated result  users can tolerate.

Based on the distribution of late arrivals, our trigger determines the minimum number of executions necessary to comply with the maximum staleness bound. To this end, we assess the staleness for each instant of time (or period of time, if there are too many instants)  and place an execution at the time that violates the bound; we iteratively repeat this process until we reach maximum allowed lateness. Due to the irregular nature of the distribution, it is likely that the staleness of the last pair of executions is smaller than all the others, meaning that the maximum staleness that we obtain, in any pair of executions, could be lower for the same amount of executions. 

To minimize and balance  staleness across pairs of consecutive executions, we apply an optimization algorithm (variation of gradient descent~\cite{DBLP:journals/corr/Ruder16}). It  minimizes the maximum staleness and returns the instants of time where we should re-execute the window. It starts with an arbitrary configuration of execution times (to optimize, we make the starting execution times correspond to the places where the distribution of late arrivals has higher relative density). After, it adjusts the execution times based on the negative gradient of  staleness in order of time. We repeat this process until  standard deviation of all staleness values is very close to zero (i.e., staleness is balanced across pairs of executions and the maximum is already at the minimum), or when a maximum number of iterations is reached, so  we can bound the time spent with this process. Due to the strategic placement of the first execution times, we found out that our algorithm converges very fast (less than a second) to the minimum value of maximum lateness, and never reached our limit of iterations.

Overall, our trigger minimizes  staleness at a minimum number of executions (necessary to achieve specified staleness bounds).

\section{Implementation}
\label{sec:impl}
We implemented \name as a state backend in Apache Flink version 1.1.1. Our source code is publicly available~\cite{flinkcode}.
We are currently engaging in transferring this technology to the Flink codebase, and have consequently initiated an issue in the Flink tracking system.
%
Next, we describe the implementation of the key aspects of our state backend.

\spara{Transparency to  applications.} To make use of our Flink backend, applications only need to specify an option in the stream environment configuration.

\spara{I/O Scheduling and Priorities.}
Using \pbuckets and \sbuckets can decouple the process of feeding window operators from the I/O activity with persistent storage.
Destaging data is carried out in the background with low priority, not to impact the performance of other operators.
In contrast, staging should have maximum priority, since  data to be fetched from the \sbucket is required immediately by the window operator that is executing.

A challenge in this context is that both staging and destaging  are I/O intensive operations and can interfere with one another.
Moreover, there are events being written simultaneously to destaged windows, which also causes I/O activity.
To prevent I/O contention, we resort to a single thread whose sole responsibility is to serialize and prioritize requests, and to perform all I/O related operations on persistent storage. This thread assigns different priorities to different operations, according to their potential impact on performance: pre-staging has maximum priority, followed by writing late events, and then destaging.

Although uncommon for sustainable workloads, these operations (namely destaging) might not finish in time.
This happens when the time between the start of the operation and when the data is needed is not sufficient to carry out the operation entirely, while possibly interleaved with other operations (e.g., destage operation being interrupted multiple times by staging requests).
If destaging is incomplete, it means that we could have released and saved more memory; if staging is incomplete, it means that operators might experience some I/O latency. However, given the priority of operations, the fact that fetching is done in the background, and the need for long term sustainability of the workloads, we believe this to be an unlikely event in practice.

\spara{Input iterator.} The input events that are accumulated in a window state are exposed to the application-specific processing functions through an iterator.
In existing implementations, iterators are initialized in an eager way: the corresponding data structure object (e.g., list) is allocated in memory with all its contents (some of which might not even be used by the window function).
Since the initialization time can be high, especially if these contents are not in memory, an eager iterator might squander memory and CPU time.
In contrast, \name uses \emph{lazy iteration}: input events are retrieved from the \sbucket as they are requested.
For example, when the iterator is called for the first time, it can issue a staging request to start staging events from the \sbucket, while at the same time it returns events from the \pbucket to the window operator.

\spara{Staging and serialization.} During destaging operations, a potentially large number of events needs  transferring from memory to persistent storage through serialization.
To speed up this CPU-intensive task, we use multiple threads serializing  blocks of events concurrently, writing out to disk in sequentially accessed files.
The blocks are the basic unit inside \pbucket.
We also use  multithreading for deserialization in staging operations.

In \name, we rely on JSON serialization since it is not the bottleneck in our experiments, and allows us to better control the file partitioning.
Using better performing serialization schemes (e.g. Kryo, Protocol Buffers, Avro) is straightforward, although orthogonal to our main goal.
The  serialization used can be easily changed in \name, e.g., to also compress data.
Note that Flink itself already ensures application data types to be serializable. 

%% file: sections/evaluation.tex
\newcommand{\myfigs}[2]{\myfigsss{#1}{#2}}

\newcommand{\myfigss}[2]{
\begin{figure}
	\centering
    \begin{subfigure}[t]{0.49\columnwidth}
		\includegraphics[width=\columnwidth]{1-#1}
        \caption{average}
        \label{fig:1-#1}
    \end{subfigure}
    \begin{subfigure}[t]{0.49\columnwidth}
    	\includegraphics[width=\columnwidth]{2-#1}  
        \caption{bigrams}
        \label{fig:2-#1}
    \end{subfigure}
    \begin{subfigure}[t]{0.49\columnwidth}
		\includegraphics[width=\columnwidth]{3-#1}
        \caption{stock market}
        \label{fig:3-#1}
    \end{subfigure}
    \begin{subfigure}[t]{0.49\columnwidth}
    	\includegraphics[width=\columnwidth]{4-#1}  
        \caption{LRB}
        \label{fig:4-#1}        
    \end{subfigure}
    \caption{#2}
    \label{fig:#1}
\end{figure}
}

\newcommand{\myfigsss}[2]{
\begin{figure*}[h!]
	\centering
    \begin{subfigure}[t]{0.245\textwidth}
		\includegraphics[width=\columnwidth]{1-#1}
        \caption{average}
        \label{fig:1-#1}
    \end{subfigure}
    \begin{subfigure}[t]{0.245\textwidth}
    	\includegraphics[width=\columnwidth]{2-#1}  
        \caption{bigrams}
        \label{fig:2-#1}
    \end{subfigure}
    \begin{subfigure}[t]{0.245\textwidth}
		\includegraphics[width=\columnwidth]{3-#1}
        \caption{stock market}
        \label{fig:3-#1}
    \end{subfigure}
    \begin{subfigure}[t]{0.245\textwidth}
    	\includegraphics[width=\columnwidth]{4-#1}  
        \caption{LRB}
        \label{fig:4-#1}        
    \end{subfigure}
    \caption{#2}
    \label{fig:#1}
\end{figure*}
}

\newcommand{\qa}{Does \name handle memory pressure effectively by offloading state to disk when needed?}
\newcommand{\qb}{What is the overhead of \name, for the case where Flink is able to operate fully in-memory?}
\newcommand{\qc}{What are the benefits of each individual optimization?}
\newcommand{\qd}{Can \name comply with maximum staleness bounds while using resources efficiently?}


In order to validate and demonstrate the effectiveness of \name we conducted an experimental evaluation of our prototype. The main objective of the evaluation was to provide answers to the following questions.

\begin{squishlist}
\item[\textbf{Q1}] \qa
\item[\textbf{Q2}] \qb
\item[\textbf{Q3}] \qc
\item[\textbf{Q4}] \qd
\end{squishlist}

\spara{Workloads.} Our experiments are based on two micro-benchmarks, \emph{average} and \emph{bigrams}, and two real-world scenarios, \emph{stock market} and \emph{Linear Road Benchmark (LRB)}. The micro-benchmarks correspond to a computation dataflow that applies a single windowed function over a data stream, calculating either the average of a stream of randomly generated integers, or all bigrams for a stream of real twitter posts (tweets). Both of these computations are non-blocking, with bigrams having much higher time complexity (2-3 orders of magnitude higher).

The first benchmark application, stock market, implements an official Flink example of a prototypical complex dataflow~\cite{stockmarket}. The application receives a (synthetic) data stream of stock market prices for different stock symbols. Over this stream, it applies rolling aggregations per stock (min, max, mean) in a sliding window (of 10 seconds every 5 seconds). Then, it uses a custom tumbling windowed function to detect when the price of a stock has suffered a variation of at least 5\%, and emit the corresponding stock symbols, as price warning alerts, to the next downstream operator. This operator, in turn, counts the price alerts per symbol in a tumbling window.
In a second substream, the  application receives a (synthetic) stream of tweets with mentions of stock symbols, and counts the number of mentions per symbol in a tumbling window. It finally joins the two substreams on key symbol, and computes the correlation between the number of symbol mentions and the number of alerts per symbol using a custom function with a tumbling window. This application must handle late events so that decision makers can rely on accurate historical information.


The second benchmark consists of a variable tolling system for a fictional expressway structure based on the Linear Road Benchmark (LRB)~\cite{Arasu:2004:LRS:1316689.1316732}. This system calculates different toll rates for different segments of a roadway based on their levels of congestion. 
The data stream that is fed as input to the dataflow is generated by the MIT-SIMLab (a simulation-based laboratory)~\cite{YANG1996113} and consists of vehicle position reports. 
The LRB dataflow can be summarized as follows. First, position reports are issued every 30 seconds by a transponder at each vehicle, identifying its exact location in the expressway system. These reports are used in two distinct substreams: 1) they are aggregated in a minute-long window to compute the number of vehicles and their average speed for every segment of every expressway; and 2) they are aggregated in a minute-long window with a custom function that detects the existence of accidents for every expressway segment. Subsequently, these two substreams are joined by key on segment; then, a custom function computes the corresponding toll based on the number of vehicles, their average speed, and the existence of accidents in a minute-long window.
Position reports might suffer temporary network disconnection or arbitrary delays, and it is necessary to incorporate the effects of late events, e.g., to ensure the accuracy of the system and of the decisions that affect billing and incentives to redirect traffic.

\spara{Event timestamps.} For all referred scenarios and windows, we assign timestamps when events are produced by  data generators. To do this, we read the current system clock and subtract a time value to make it fall either in the current window or in a past window, thereby simulating event delays:

\vspace{1pt}
\noindent $ ts = currentTime - windowIndex \times windowDuration $
\vspace{1pt}

Thus, the timestamp \emph{ts} is given by the current time subtracted by a certain number (\emph{windowIndex}) of window lengths. To simulate a realistic delay, we set the \emph{windowIndex} based on a log-normal distribution (mean and stddev are 0 and 1 respectively). Thus, the likelihood that a window receives an event decreases exponentially, as expected in most practical scenarios.

\spara{Setup.} For all experiments, we compared the use of our backend, \name, with a baseline consisting of Flink's existing backend, whose implementation is only able to retain all window state in memory. Note that our gains and overheads come from the custom (i.e., user-defined) windowed functions; for other stateful operators, that do not rely on ListState, we perform similarly to baseline. For our backend, we used the standard policy (see~\S\ref{sect:architecture}) throughout the experiments. Furthermore, we set the \pbucket size to $500,000$, since this value is large enough to hold all events within a window in memory (before events are destaged).

After the system reached steady state (taking at least 10 watermarks), each execution had the duration of 30 watermarks, with an interval between watermark generation equal to the window duration. We varied the number of past windows throughout the experiments. JVM max heap was set to 8 GB.

For each workload, we set the following parameters: maximum ingestion rate, the duration of the window, and the size of the additional payload added to incoming events. The values we used for these parameters are given in Table~\ref{tab:workloads}.


\begin{table}
\begin{tabular}{l r r r}
\toprule
\multicolumn{1}{p{8mm}}{\centering\footnotesize{Scenario}} & 
\multicolumn{1}{p{18mm}}{\centering \footnotesize{Max ingestion rate (events/s)}} & \multicolumn{1}{p{15mm}}{\centering \footnotesize{Window duration (s)}} &
\multicolumn{1}{p{15mm}}{\centering \footnotesize{Payload size (bytes)}} \\ 
\midrule
  \footnotesize{Average} & \footnotesize{10000} & \footnotesize{20} & \footnotesize{2304} \\ 
  \footnotesize{Bigrams} & \footnotesize{5000} & \footnotesize{30} & \footnotesize{3584} \\
  \footnotesize{Stock market} & \footnotesize{10000} & \footnotesize{30} & \footnotesize{1664} \\
  \footnotesize{LRB} & \footnotesize{10000} & \footnotesize{60} & \footnotesize{1536} \\
\bottomrule
\vspace{2mm}
\end{tabular}
\caption{Workload parameters}
\label{tab:workloads}
\end{table}

\spara{Setting.} All tests were conducted using two machines with an Intel Core i7-2600K CPU at 3.40GHz, 11926MB of RAM memory, and HDD 7200RPM SATA 6Gb/s 32MB cache, connected by 1 Gbps LAN. One machine was used to run the data generators and the other to execute the streaming applications. This setting shows the benefits of \name on a per-node basis. 
The running environment consisted of Ubuntu 14.04.1 LTS (GNU/Linux 3.13.0-116-generic x86\_64), Java HotSpot(TM) 1.8.0\_77 and Flink 1.1.1. Our source code and the setup for the experiments is publicly available~\cite{workloadscode}.

\mpara{Q1. \qa}\newline

\myfigs{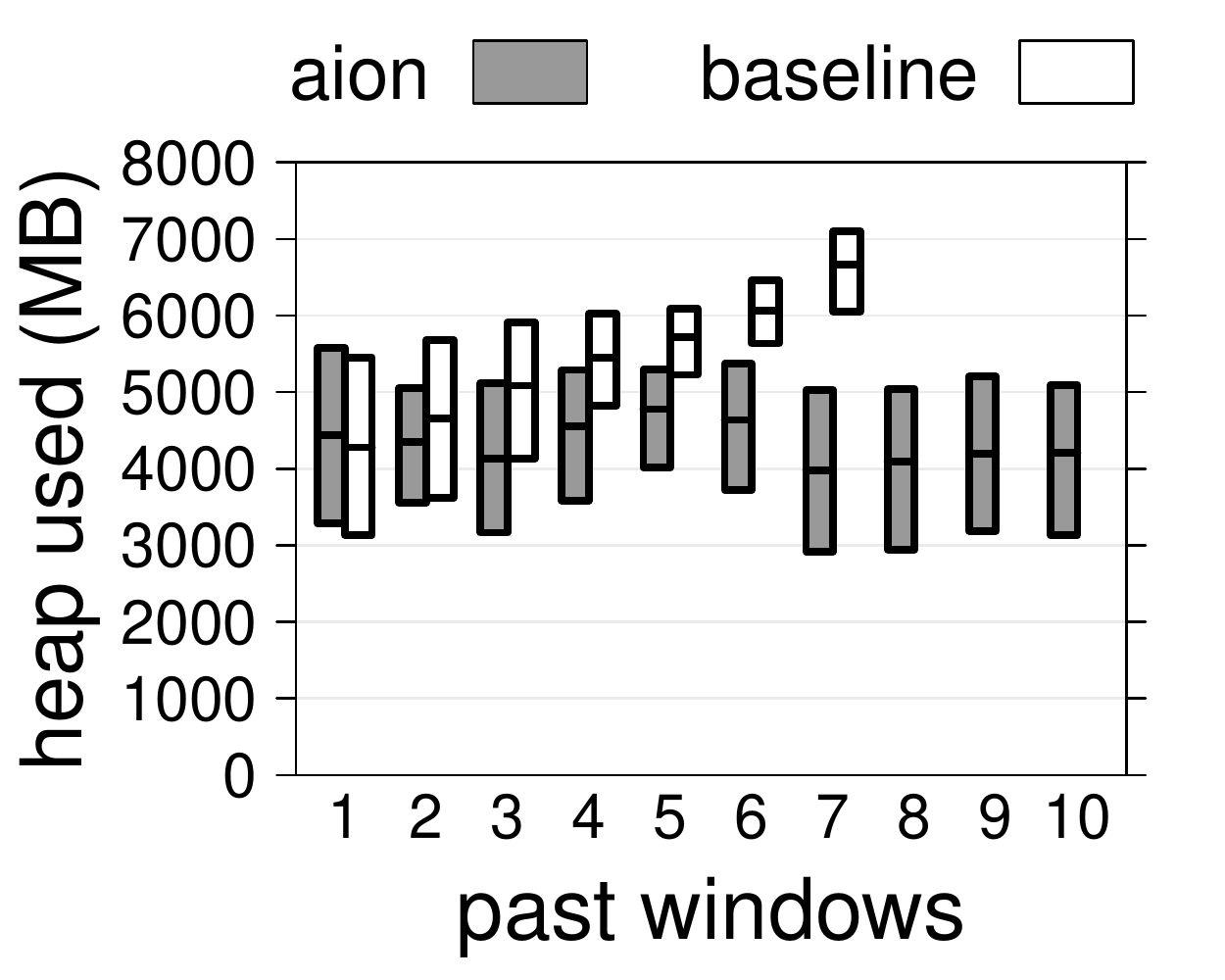}{Heap usage across  workloads. Boxes represent:  median of the data, lower and upper quartiles (25\%,75\%)}
Figure~\ref{fig:heap} shows the heap usage of our approach compared to the baseline. For the baseline system, as the number of past windows (i.e., the maximum allowed lateness of events) increases, the heap usage also increases, since more state data has to be maintained and accumulated in memory over time. The heap utilization of the baseline eventually becomes so large that the system crashes due to insufficient available memory. This happens after 7, 9, 5, and 8 past windows for average, bigrams, stock market, and LRB, respectively. In contrast, \name is able to maintain a stable and efficient memory utilization over time, roughly 3-4 GB as the median, regardless of the number of past windows, thus scaling window state for a potentially unbounded time frame. Such capability comes from the fact that \name keeps only the state of active windows in memory; past window state is destaged and kept in persistent storage. Thanks to \caching, this comes without impacting the ingestion rate (as we will see next).
Finally, \name offers significant savings in terms of median heap memory usage: it uses between 50\% (bigrams) and 24\% (stock market) less memory than the baseline.

\mpara{Q2. \qb}\newline

We now measure the overhead of \name in terms of ingestion and processing rates.
In Flink, the processing rate (events processed per second in a window) can affect the ingestion rate (events received per second), and therefore it is possible that, over time, the latter  does not remain stable at its maximum (as shown in  experimental setup). 
\myfigs{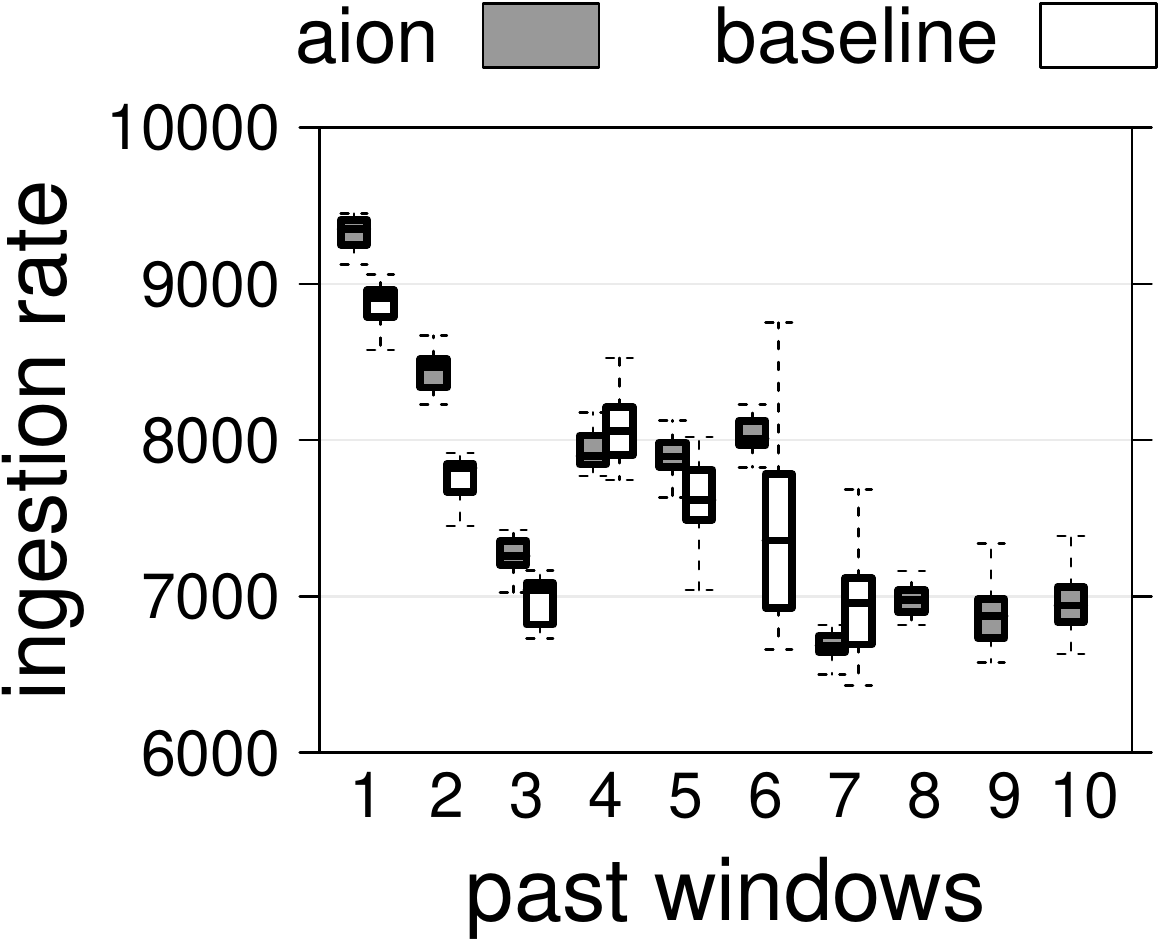}{Ingestion rate of normal (non-late) events only. \emph{Whiskers} show minimum and maximum values}

\spara{Ingestion rates.}
The ingestion rate measures the end-to-end throughput of the system, and is the most important metric for the performance of an SPS. A high ingestion rate shows that the system can keep up with its inputs.
In particular, if the time it takes to process a window exceeds the window interval, which is the risk a system incurs when offloading window state to the disk like \name, then the ingestion rate drops. Our evaluation shows that \name has virtually no impact on the ingestion rate, thanks to its use of \caching.

Figure~\ref{fig:thr} shows, for each benchmark, the ingestion rate of normal (non-late) events only. We can observe that 1) with the exception of bigrams, there are no large variations across executions for different values of the number of past windows; and 2) the differences between \name and baseline are relatively small. The higher variation in bigrams is linked to the fact that its input events (tweets) have a more variable size. Different input event sizes in bigrams, which is a computationally complex workload, cause different compute times over the execution timespan (note that processing time makes the ingestion of new events to stall in Flink).

The results indicate that \name is on par with baseline in terms of end-to-end performance: in the most favorable case, as baseline starts thrashing and crashing, \name ingested up to 17\% more events than baseline with LRB; in the least favorable case, baseline ingested up to 18\% more events than \name with LRB. All other workloads show variances between 4\% and 10\%.

\myfigs{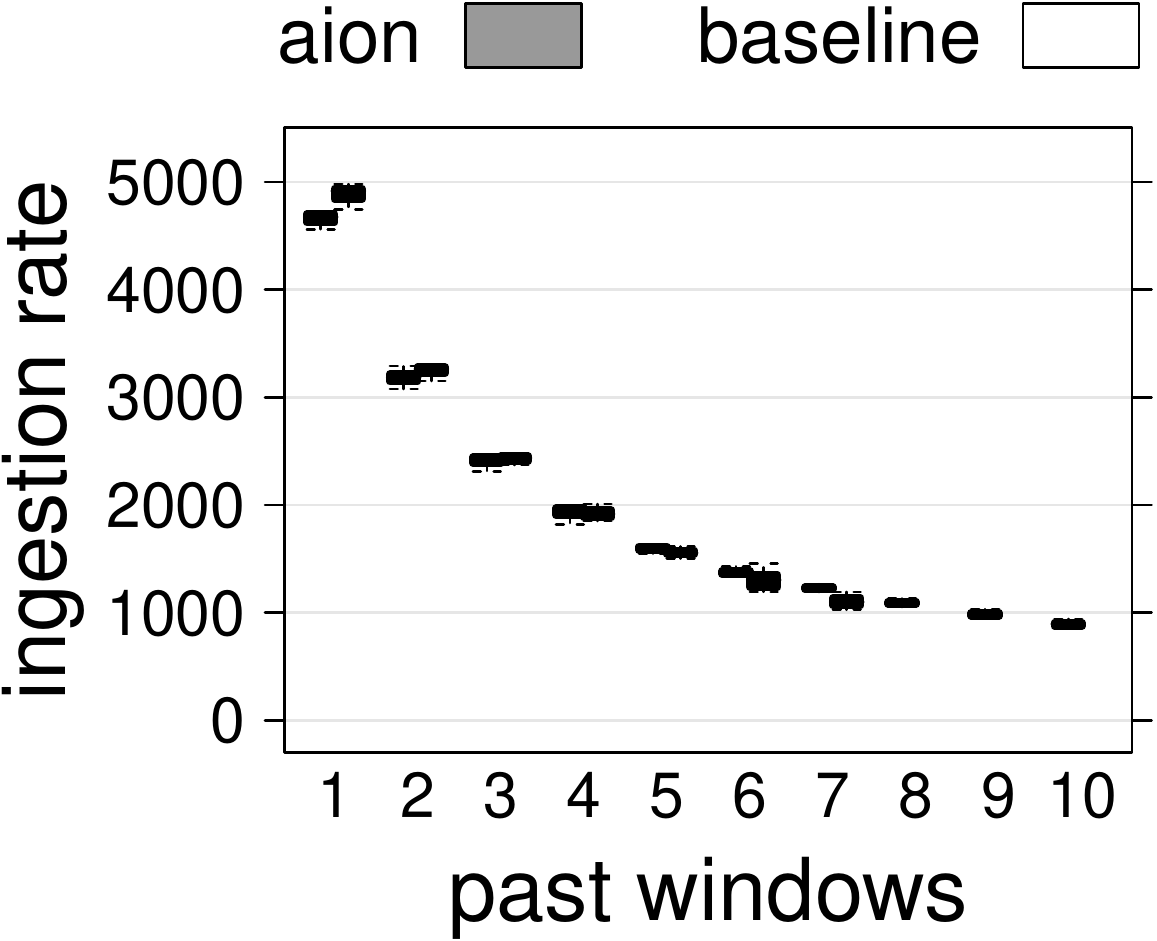}{Ingestion rate of normal and late events}

Figure~\ref{fig:thr-late} shows the ingestion rate of each benchmark for different lateness values, but this time including also late events. The ingestion rate decreases as the number of past windows increases in this case. This comes as no surprise: as a window ages, it is likely to receive fewer events, and therefore the overall ingestion rate tends to decrease as we extend the lateness timespan.
Nevertheless, the ingestion rate values for \name get slightly closer to the baseline values: we go from a gain of 12\% using \name (average) to a gain of 16\% using the baseline (stock market). Other workloads exhibited a variation ranging from 1 to 9\%. This happens because the number of normal and late events received over time decreases exponentially, which makes the differences smaller and more stable. 

\myfigs{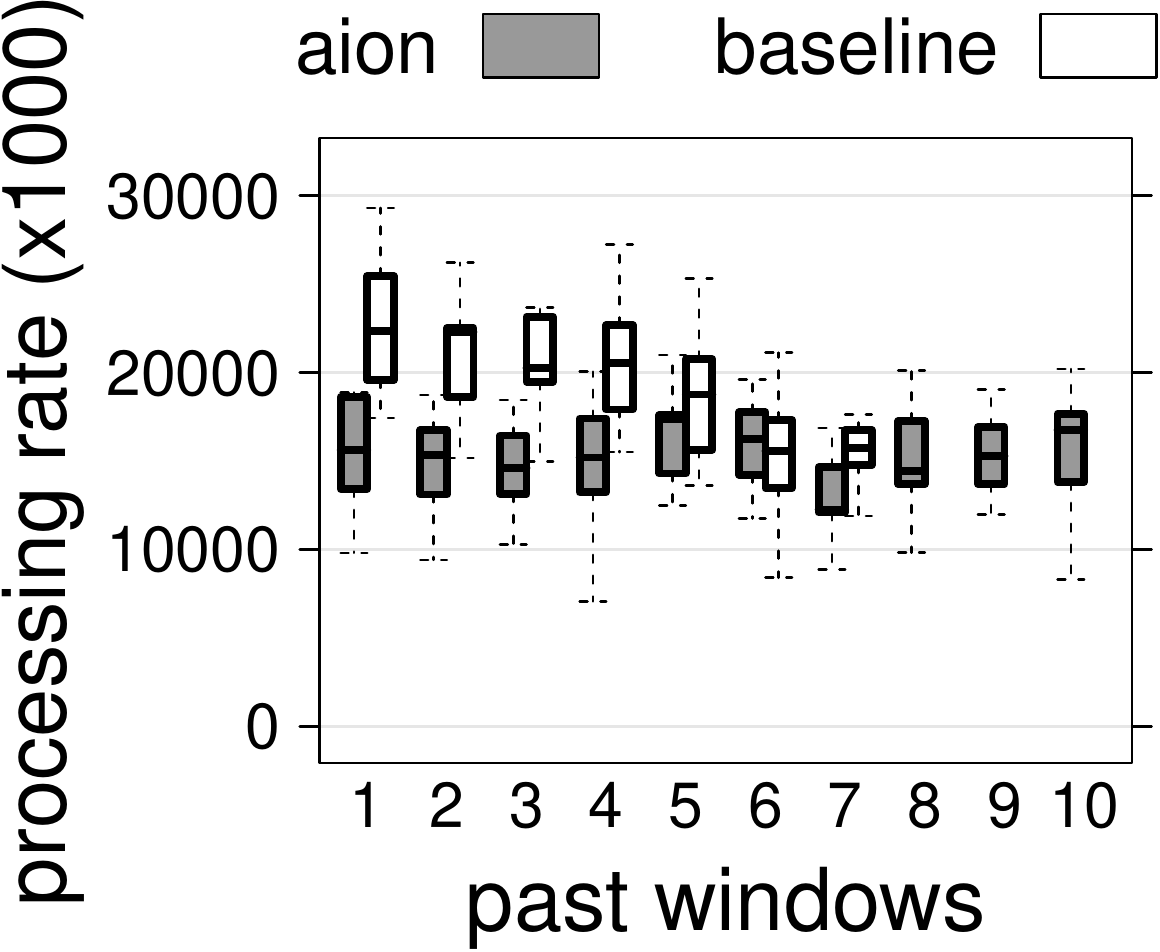}{Processing rate of normal (non-late) events only.}

\spara{Processing rate.}
The processing rate gives a more low-level insight on the overhead of \name. The previous experiments show that, in all cases, the processing rate of \name is sufficient to keep up with the ingestion rate of the application. The following experiments show that with windowing functions having high computational complexity, \name has a similar processing rate as the baseline, since the cost of fetching data from disk can be amortized (thanks to \caching). For functions with low computational complexity, \name has a relatively lower processing rate, but this does not matter in absolute terms since windows can be nonetheless computed quickly enough.

\begin{figure}
	\centering
    \begin{subfigure}[t]{0.49\columnwidth}
		\includegraphics[width=\columnwidth]{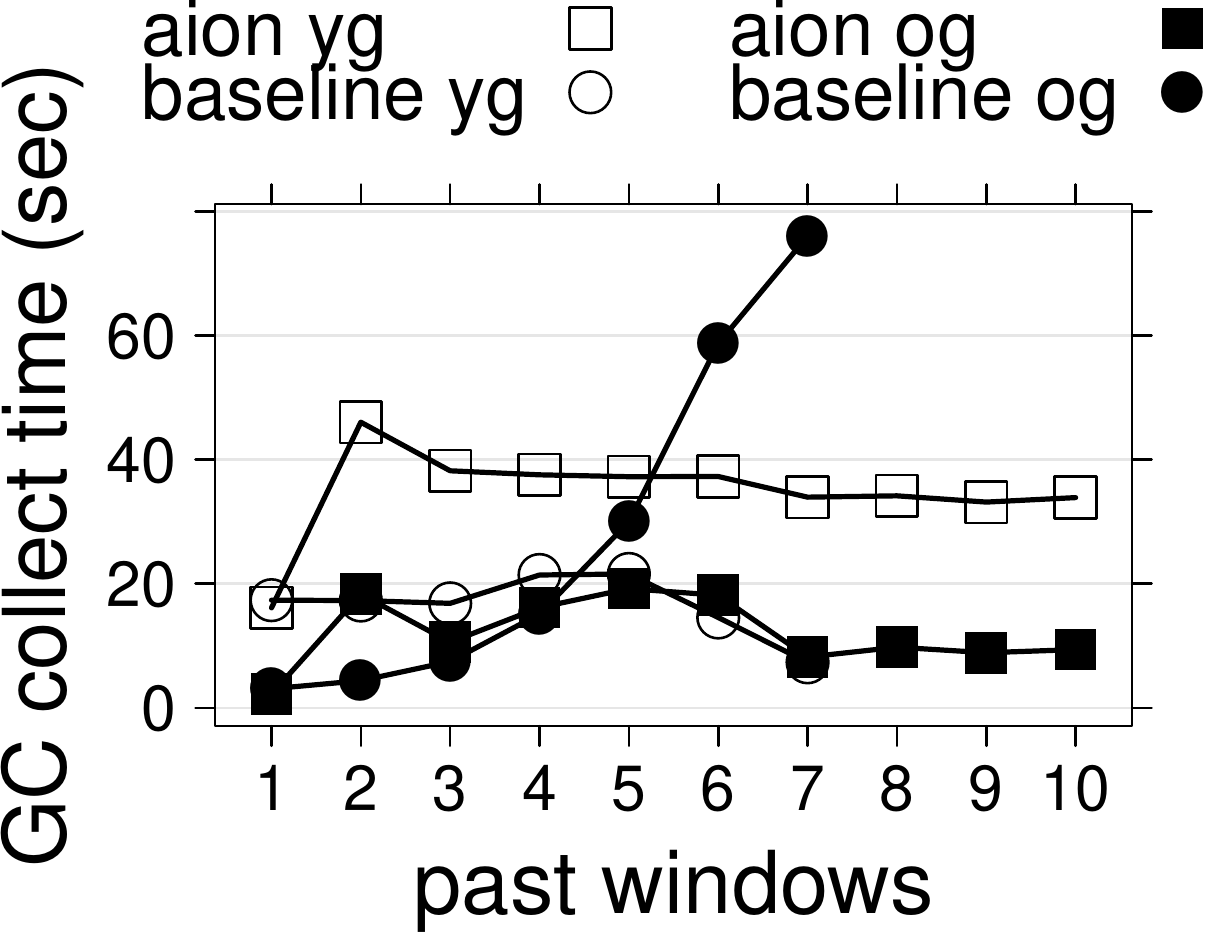}
        \caption{average}
        \label{fig:1-gc-time}
    \end{subfigure}
    \begin{subfigure}[t]{0.49\columnwidth}
    	\includegraphics[width=\columnwidth]{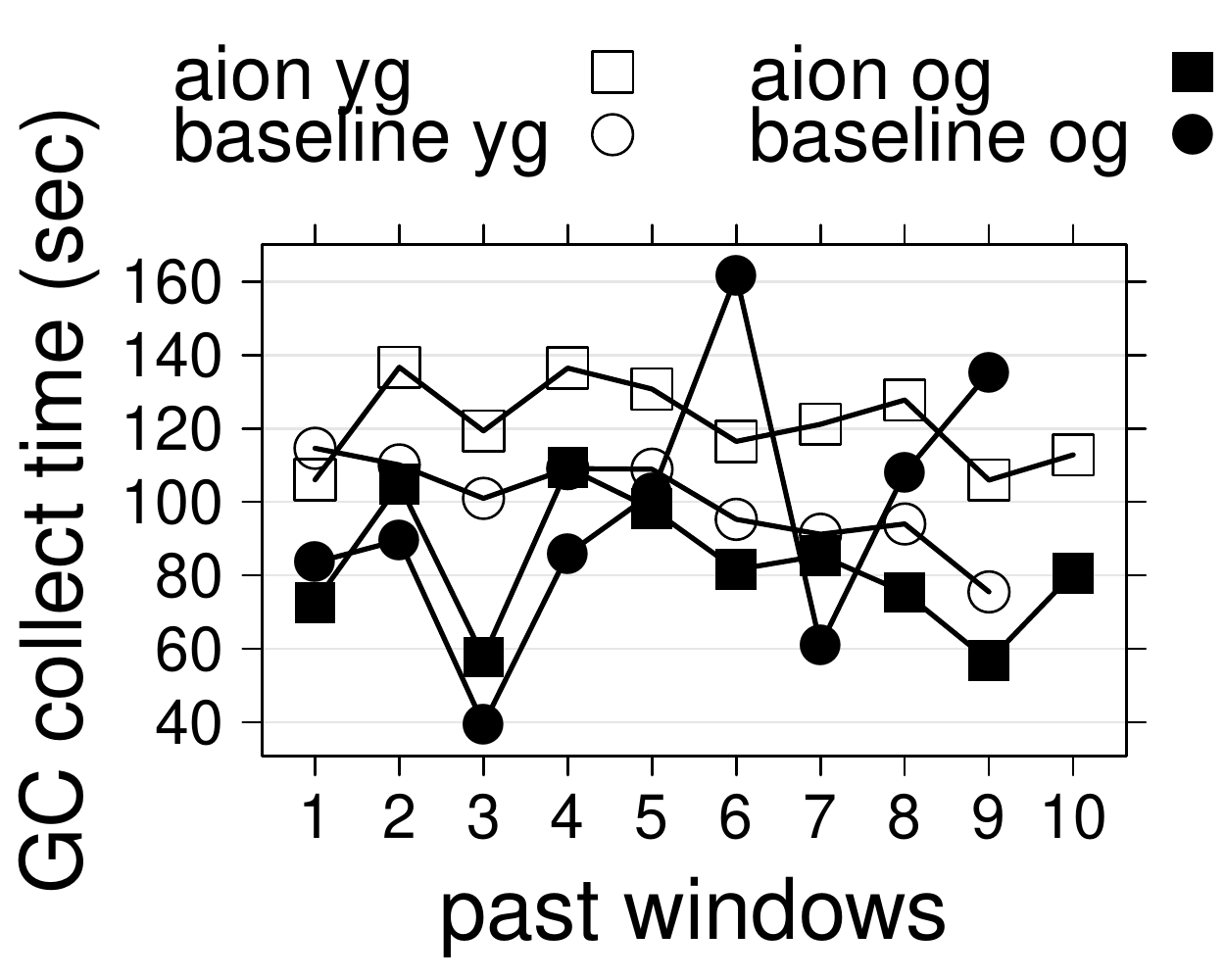}  
        \caption{bigrams}
        \label{fig:2-gc-time}
    \end{subfigure}
    \caption{GC collecting time}
    \label{fig:gc-time}
\end{figure}

Figure~\ref{fig:proc-rate} depicts, for each benchmark, the normal (non late) event processing rate of \name and the baseline, when varying between 1 and 10 past windows. Several things can be observed. First, we can see that for average, bigrams, and LRB, the processing rate of \name is mostly stable as the number of past windows increases; in contrast, the processing rate of the baseline is mostly unstable for all the considered scenarios. Second, for average and bigrams, although the processing rate of the baseline starts by being higher than \name, it follows a decreasing tendency as the number of past windows increases. This phenomenon occurs because the system starts thrashing: as the heap usage reaches close to its limit, the JVM Garbage Collector is activated for longer intervals in the old generation (as shown in Figure~\ref{fig:1-gc-time}), which in practice steals CPU time from the applications. Moreover, average exhibits more accentuated differences between \name and baseline (up to 31\%). Such differences come as a result of \name having significant more GC activity on the young generation than the baseline (as shown in Figure~\ref{fig:1-gc-time}). The increased activity is due to the additional backend data structures that we manage.


Although stock market generates a complex dataflow in terms of its streaming graph, the windowed functions themselves have a low time complexity: each window takes less than one second to process tens of thousands of events. As such, the fluctuation in the time for processing a single event is much higher. Nonetheless, because the computation time is so short, \name can still keep up with the ingestion rate, so this relative difference is not relevant in terms of end-to-end performance.

Finally, for LRB, the processing rate has less variance than stock market because the computation time is higher. When the baseline starts thrashing (after just 5 past windows), the first quartile of the processing rate drops drastically. This behavior results from alternating between high compute time (which includes GC time)
with low ingestion rate: as the GC activity for the old generation increases, processing time increases, and ingestion rate decreases; as such, for the next watermark, fewer events are expected, which makes GC activity and processing time decrease; in turn, this makes ingestion time increase and this cycle repeats. Furthermore, the relative difference becomes between \name and baseline becomes significant because this workload has two memory intensive custom functions, which results in a higher GC activity on the young generation.

\myfigs{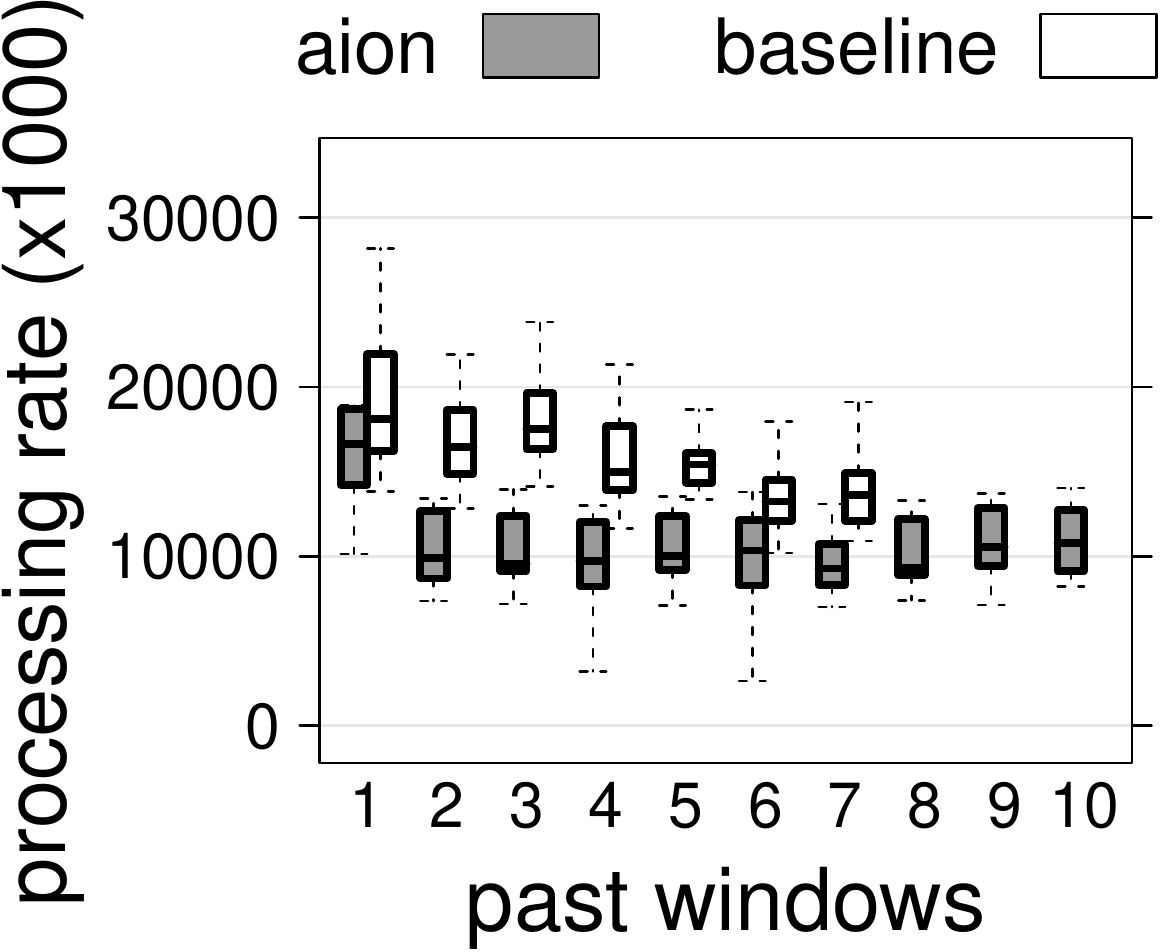}{Processing rate of normal and late events}


Figure~\ref{fig:proc-rate-late} shows the processing rate when late events are included. Variance is generally reduced, especially for stock market and LRB. Similarly to what was described before for Figure~\ref{fig:thr-late}, the number of events is greatly reduced as a window gets older, and this attenuates the differences between \name and baseline over time. 

To summarize, there are two main take-aways. First, \name overheads are realistically low, as the higher the complexity of the custom windowed functions, the closer is \name processing rate to the baseline.
Second, in addition, the processing rate only becomes relevant as an overhead when it makes the streaming application not sustainable across time. As long as the system is able to continuously provide results for every fixed time interval (sustainability condition), corresponding to the latency requirements defined through window duration values, the processing time overhead can be disregarded.

\mpara{Q3. \qc}\newline

We now assess the impact (contribution and relevance) of the individual optimizations: pre-staging, multi-threading serialization, and single thread (sequential) I/O. We employ the average workload, since it is a simple pipeline with single window  and  low complexity function (i.e., where optimization effects are more isolated and events need to be fetched quicker). The optimizations are especially important to reduce the fetching time of a window operator when most of the state data resides in the \sbucket, which is the case with the standard policy when the allowed lateness time expires.



\newcommand{\aion}{aion-full\xspace}
\newcommand{\stage}{no-pre-stgng\xspace}
\newcommand{\serial}{no-mt-srlz\xspace}
\newcommand{\io}{no-sqntl-io\xspace}

\begin{figure*}[h!]
	\centering
    \begin{subfigure}[t]{0.32\textwidth}
      \includegraphics[width=0.98\columnwidth]{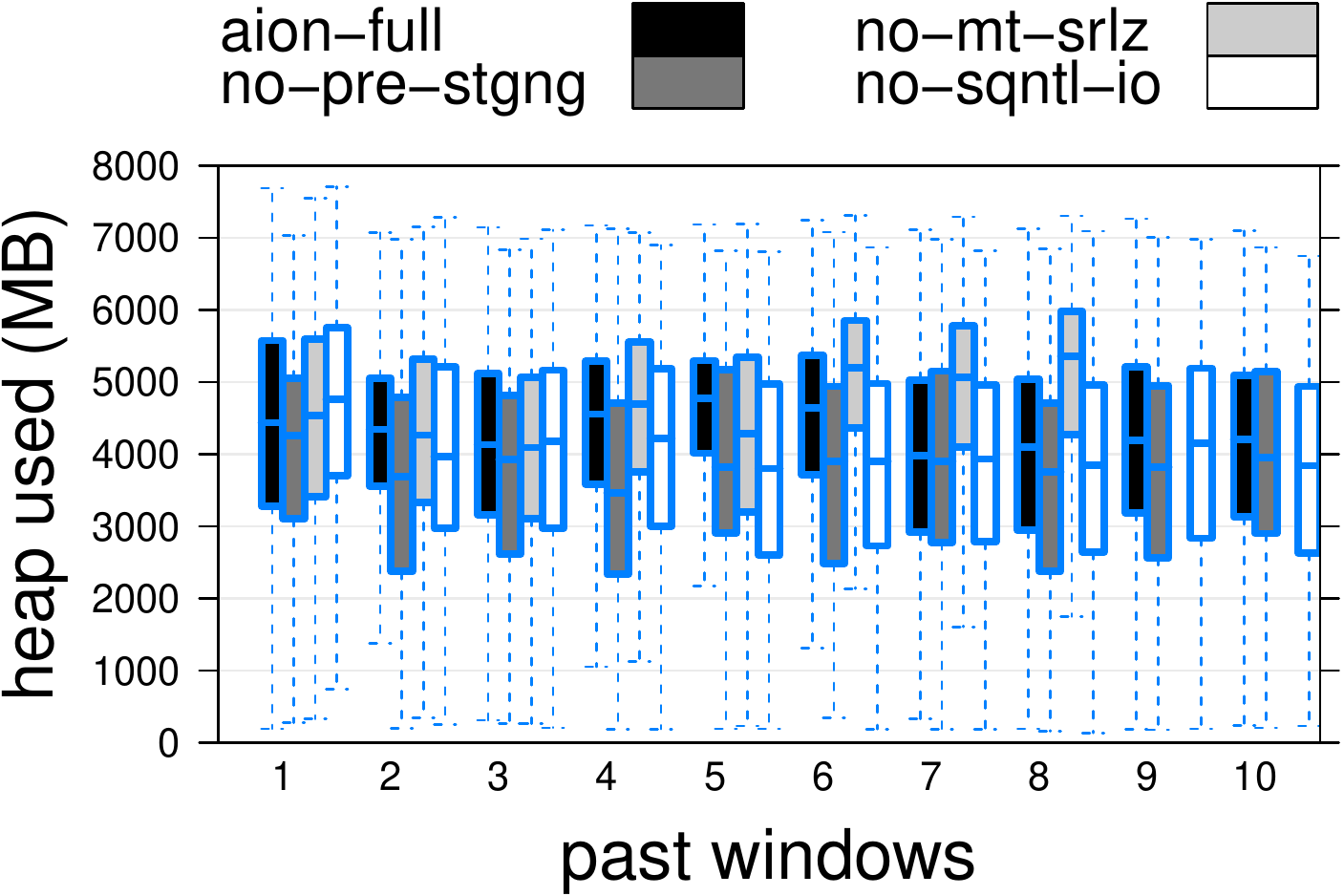}
      \label{fig:optimization-heap}
    \end{subfigure}    
    \begin{subfigure}[t]{0.32\textwidth}  
      \includegraphics[width=0.98\columnwidth]{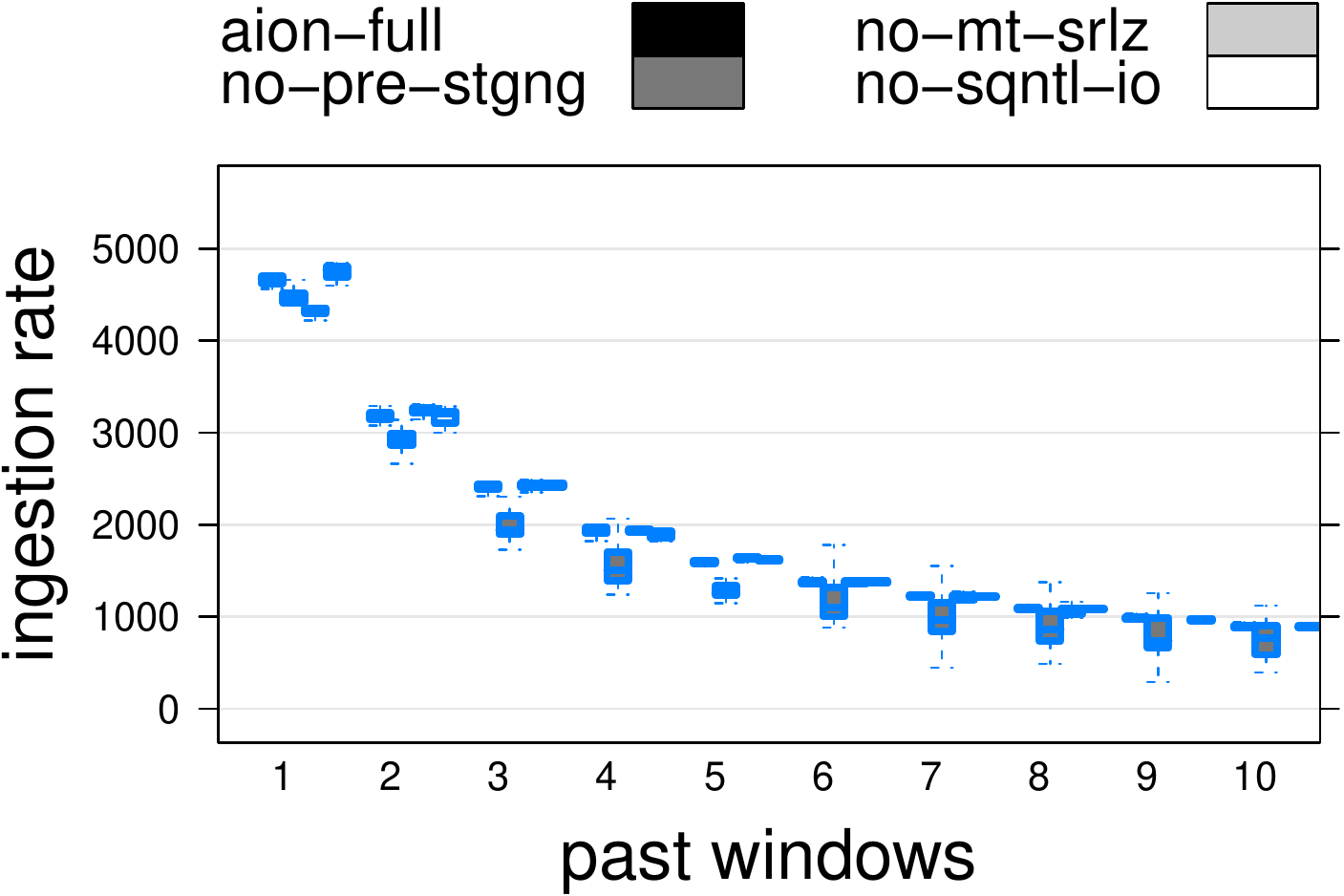}
      \label{fig:optimization-thr}    
    \end{subfigure}
	\begin{subfigure}[t]{0.32\textwidth}  
      \includegraphics[width=0.98\columnwidth]{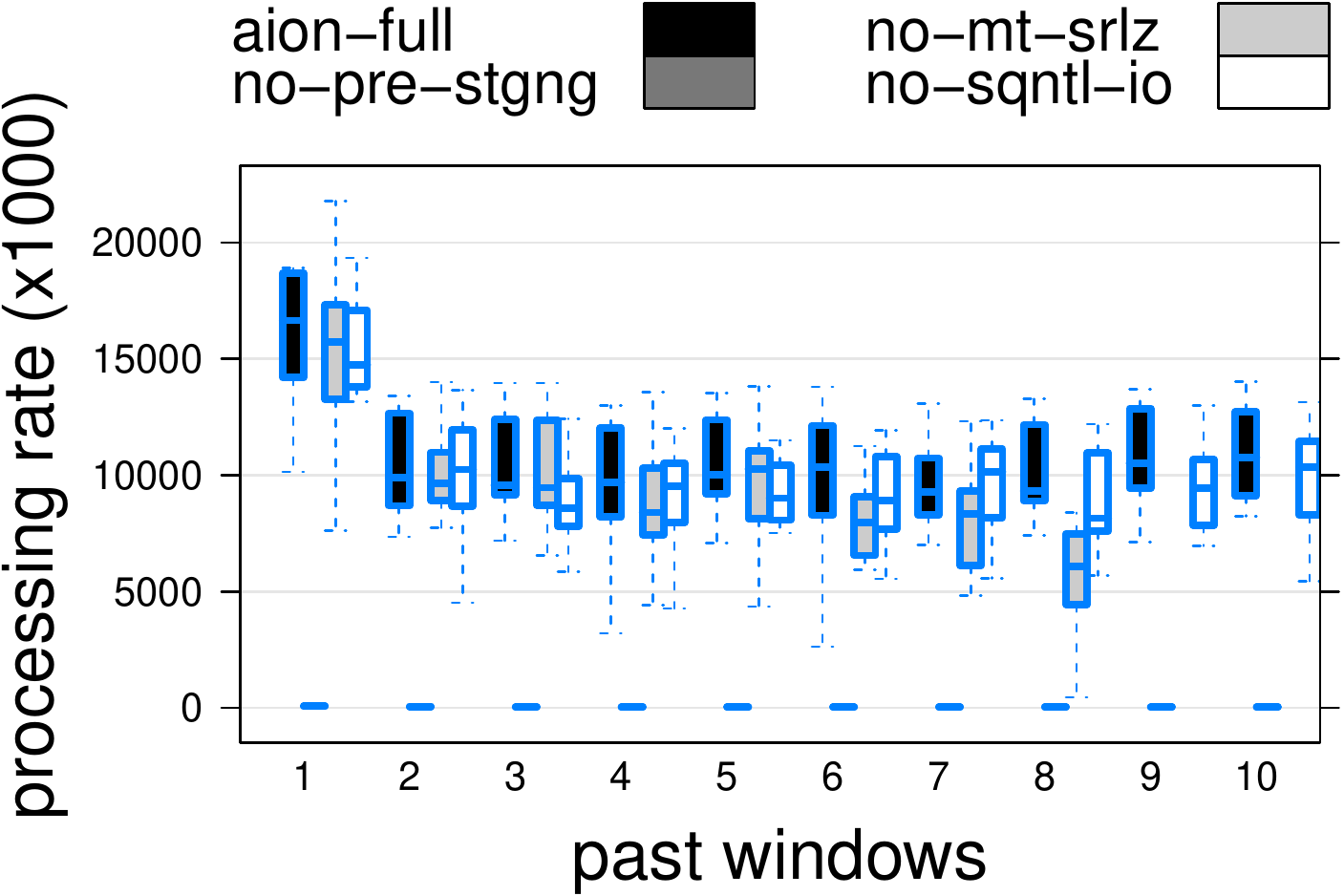}
      \label{fig:optimization-proc-rate-late}    
    \end{subfigure}
    \caption{Effect of optimizations on normal and late events for the average workload. \emph{\aion} corresponds to the system fully optimized (with pre-staging, multi-thread serialization, and single I/O thread); \emph{\stage} is \name with pre-staging off; \emph{\serial} is \name with single serialization thread; and \emph{\io} is \name with multi-threads performing I/O operations simultaneously.}
    \label{fig:1-optimization}
\end{figure*}

Figure~\ref{fig:1-optimization} shows the effect that each optimization has on the heap usage, ingestion and processing rate of all events, when varying the number of past windows from 1 to 10. First, for \stage, we can see that it uses less memory for average than \aion (left sub-figure), which is natural since pre-staging loads state data in advance, and thus keeps memory occupied for a slightly longer time. However, \stage performs significantly worse -- by 2 orders of magnitude -- in processing rate (bars close to zero in right sub-figure), since it fully exposes the I/O latency by accessing the persistent storage (\sbucket) while the function is executing. As a consequence of the longer processing times with \stage, the corresponding ingestion rate (central sub-figure) is also affected negatively: \aion receives roughly 20\% more events for average. We can thus conclude that pre-staging is a key feature in \name.

Second, we may observe that \serial is not able to stabilize heap usage as the lateness time increases, ending up crashing after 8 past windows. This shows that a single thread for serialization is not sufficient to serialize data fast enough in destaging operations, leading thus to poorer memory savings. Similarly, one thread for deserialization is also not enough, since the processing rate falls as the number of past windows increases and more events have to be staged.

Finally, we can infer, for all of three metrics, that the performance values are closer between \io and \aion, yet \io reveals a decreasing trend  in processing rate as the lateness time increases. This trend  results from the fact that as we keep more events from the past in memory, the more likely it is to have destaging and staging operations to be incomplete at the time when the window execution starts. Staging operations, which impact the processing rate, should have higher priority on completion than destaging operations (i.e., memory savings are not as critical as complying with latency requirements), and that is what \name achieves with a single thread that prioritizes I/O operations.

\mpara{Q4. \qd}\newline

\begin{figure*}[h!]
  \centering
  \begin{subfigure}[c]{0.4\textwidth}
    \includegraphics[width=\textwidth]{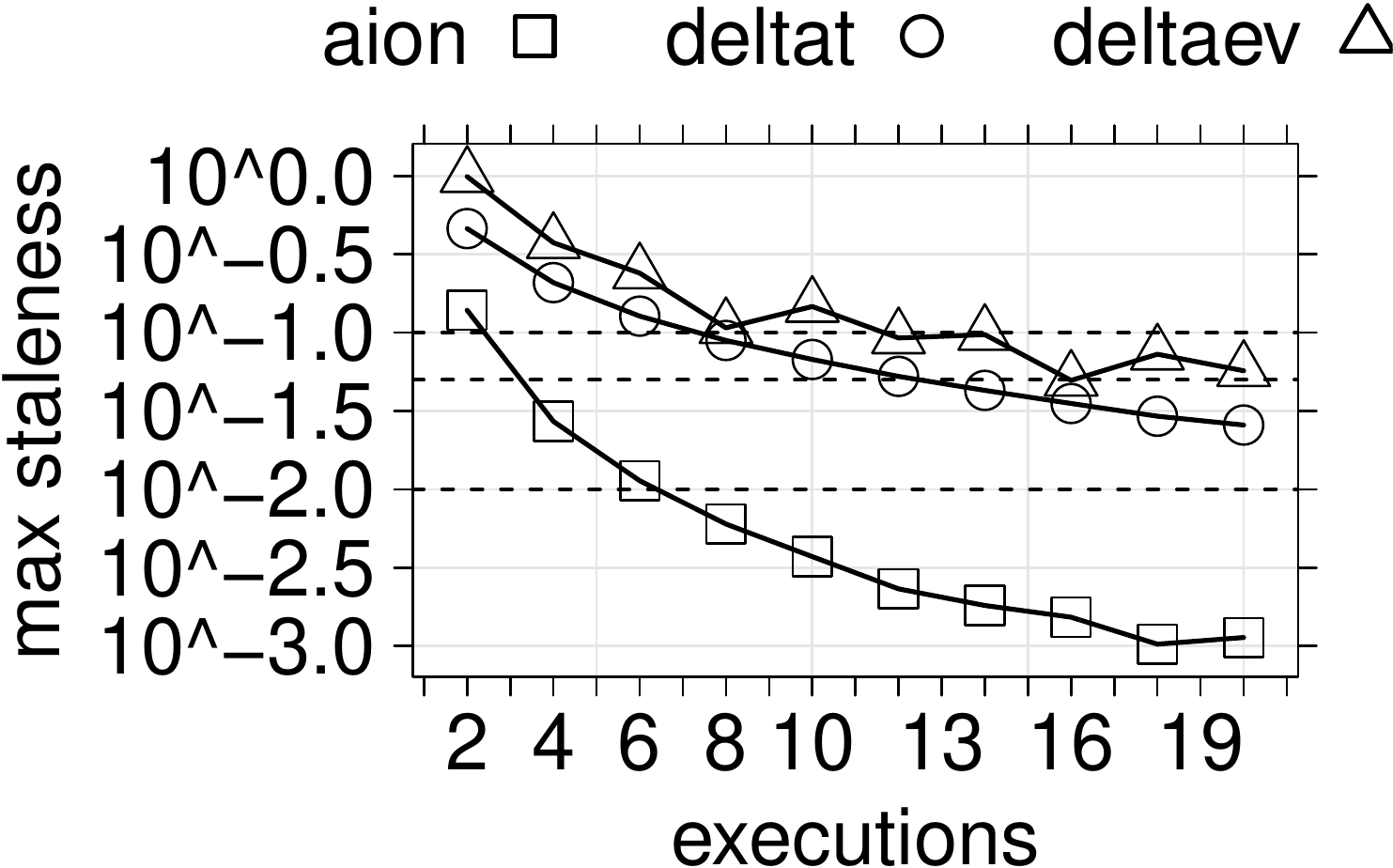}
  \end{subfigure}
  \begin{subfigure}[c]{0.4\textwidth}
    \includegraphics[width=\textwidth]{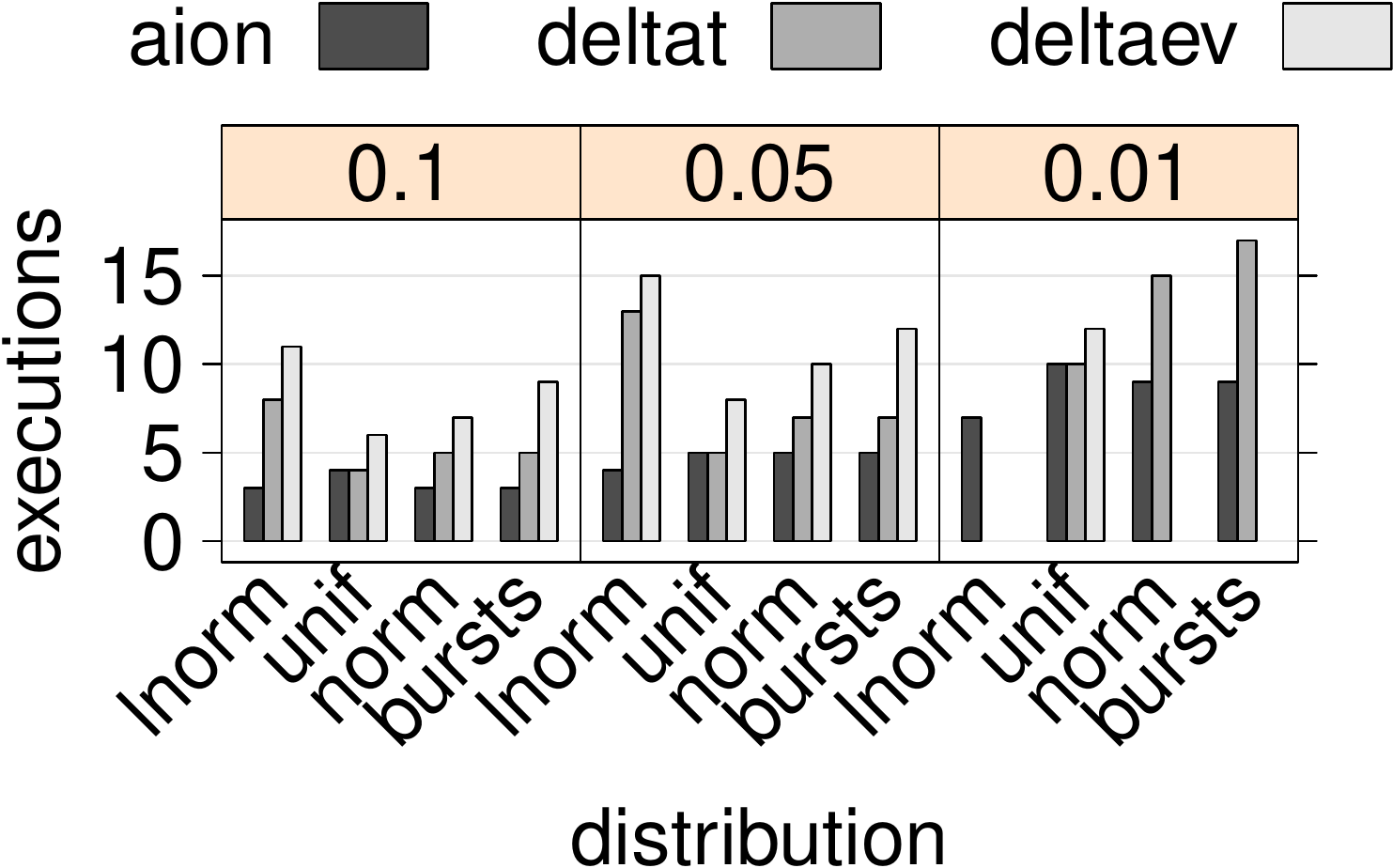}
  \end{subfigure}
  \caption{Maximum staleness, in logarithmic scale, for varying no. of executions for the log-normal distribution (left-side); minimum no. of executions necessary to reach a staleness bound that is 10, 5 and 1\% of the maximum allowed lateness time, for different distributions (right-side).} 
  \label{fig:trigger}
\end{figure*}
We now assess the effectiveness of our trigger described in \S\ref{sect:trigger}. Here we do not resort to the standard policy anymore, which was used in previous experiments to assure a fair comparison against baseline. For late windows, our trigger attempts to minimize the maximum staleness among pairs of execution times while not using more than the necessary resources to comply with user-defined staleness limits.

For a log-normal distribution of late events, the left side of Figure~\ref{fig:trigger} depicts the the maximum staleness obtained across executions for different triggers. \name is our trigger, \emph{deltat} corresponds to a punctuated trigger that executes periodically at every time interval, and \emph{deltaev} is a trigger that executes at every $x$ events, where $x$ is the total number of events expected divided by the number of executions. 
Our trigger achieves increasingly lower maximum staleness in relation to the standard triggers \emph{deltat} and \emph{deltaev} for the same amount of executions. Moreover, the standard triggers take more executions than \name to reach the bounds of $0.1$ and $0.05$, and fail to reach the bound of $0.01$ within 20 executions.

The right side of Figure~\ref{fig:trigger} shows that our trigger is also effective for other distributions of late events. Apart from the log-normal $(lnorm)$, we considered $(unif)$ a distribution that makes late events uniformly distributed across time; $(norm)$ a normal distribution of events; and $(bursts)$ a mix of normal distributions that generate bursts of late events. We show for each distribution what is the minimum number of executions to reach the considered bounds ($0.1, 0.05, 0.01$).

The \emph{deltat} trigger is as good as \name for the \emph{unif}, since it places the executions uniformly distributed in time, following the same trend of late event arrival. However, a uniform distribution is not realistic: the arrival of late events tends to have a more irregular behavior (due to temporary disconnected devices, network delays, etc.). For the other distributions, \name reached all the bounds with less executions than those of standard triggers. The major gain was for the log-normal distribution with a bound of $0.05$, where \name performed only 31 and 27\% of the executions of \emph{deltat} and \emph{deltaev} respectively. Moreover, standard triggers failed to reach the small bound of $0.01$ for \emph{lnorm} within 30 executions. This means that \name is able to comply with small staleness bounds at the minimum possible number of executions.

%% file: sections/related-work.tex



Stream processing has been researched for some time~\cite{Stephens1997}. Despite its maturity, there has been a recent surge in interest, mainly due to necessity of processing large amounts of data in real-time~\cite{storm}. SPSs that operate with a clear semantic of event-time with emphasis on correctness have emerged only in the last few years~\cite{dataflow,murray2013naiad}. Even more recently, modern SPSs started acknowledging dealing with lateness, such as Google Cloud Dataflow~\cite{dataflow} and its predecessor Millwheel~\cite{akidau2013millwheel} that refer to the difficulty of picking a maximum allowed lateness, yet always leaving the task to the developer.

State spilling has been proposed to handle memory overloaded operators by transferring parts of the state from memory to disk~\cite{Liu:2006:ROS:1142473.1142513}. This state spilling is limited to non-window operators, despite authors acknowledging that tackling window constraints would require interleaving in-memory execution with disk management, and would bring a new set of challenges, such as the timing of spill, timing of clean-up, and selection of data to clean-up. Our work addresses these challenges, that have remained unresolved until now~\cite{to2017survey}. Particularly, we offer a comprehensive solution to deal with the problematic of lateness, where we go beyond a solution that simply spills data to disk naively. We manage state across memory and disk with \caching, avoiding processing rate penalty due to I/O overhead, and \gc, releasing resources when they are estimated as not needed anymore. Also, we offer a trigger that minimizes staleness while using resources efficiently.

When broadening the scope of the comparison to other types of systems, a few have addressed the issue of handling late events.
One such example is Photon~\cite{Ananthanarayanan:2013:PFS:2463676.2465272}, which is a system deployed at Google for joining the click-stream with ads, provides exactly-once semantics on unordered streams, coupled with robust fault tolerance.
The design of Photon is quite different from the stream processing engines we are considering (e.g., it uses Paxos), since it is a specific solution developed for a few critical applications. In particular, it is not clear how their solutions would apply to existing distributed stream processing systems.

Another example is Samza~\cite{Noghabi:2017:SSS:3137765.3137770}, a stream processing system created at Linkedin. Without scaling horizontally with more containers, Samza acknowledges that disk spilling is necessary in order to scale to large state, however the authors refer to this as an orthogonal problem to their approach and do not provide a concrete solution.

Li et al.~\cite{li2008out} argue that setting an appropriate maximum lateness (referred as slack) is extremely difficult in practice. Therefore, they propose out-of-order processing, together with stream punctuation for watermarking, as a solution.
However, the design and implementation of the watermarking scheme are not discussed in detail, and late events are never considered. In contrast, our proposal presents the design and implementation of a complete solution in the context of real-world, non-ideal watermarks and late events.

Finally, fault-tolerance and checkpointing are related, but orthogonal topics: tolerating machine failures may be done by storing state in a persistent medium; however, the solutions that are used for tolerating faults do not necessarily apply to the problem tackled in this paper (e.g., such solutions do not involve offloading state from memory). This is the case, in particular, for the solution used by Flink~\cite{Carbone:2017:SMA:3137765.3137777,carbone2015lightweight}.

%% file: sections/conclusion.tex
This paper presented \name, a comprehensive solution to deal with late events, tailored to memory-intensive long-lived windows with potentially large periods of tolerated lateness. First, \name offloads window state from memory to disk and recovers it through \caching at strategic times. Second, \name estimates the best maximum allowed lateness based on the continuous observation of the distribution of late events over time (\gc). Finally, \name provides a customized trigger for past windows that is able to determine the execution times that minimize result staleness at a minimum amount of executions (necessary to comply with user-specified staleness bounds).

Experimental evaluation indicates that \name is capable of maintaining sustainable levels of memory utilization while still preserving high throughput, low latency, and low staleness.






%% file: main.bbl
\begin{thebibliography}{10}

\bibitem{workloadscode}
{Aion benchmarks}.
\newblock \url{https://github.com/sesteves/aion-benchmarks}.
\newblock Accessed: Feb 2018.

\bibitem{apex}
{Apache Apex}.
\newblock \url{http://apex.apache.org/}.
\newblock Accessed: Feb 2018.

\bibitem{samza}
{Apache Samza}.
\newblock \url{http://samza.apache.org/}.
\newblock Accessed: Feb 2018.

\bibitem{spark}
{Apache Spark}.
\newblock \url{http://spark.apache.org/}.
\newblock Accessed: Feb 2018.

\bibitem{storm}
{Apache Storm}.
\newblock \url{http://storm.apache.org/}.
\newblock Accessed: Feb 2018.

\bibitem{flinkcode}
{Flink 1.1.1 with \name backend}.
\newblock \url{https://github.com/sesteves/flink}.
\newblock Accessed: Feb 2018.

\bibitem{googleclouddataflow}
{Google Cloud Dataflow}.
\newblock \url{https://cloud.google.com/dataflow/}.
\newblock Accessed: Feb 2018.

\bibitem{stockmarket}
{Stock market example}.
\newblock
  \url{https://flink.apache.org/news/2015/02/09/streaming-example.html}.
\newblock Accessed: Oct 2017.

\bibitem{akidau2013millwheel}
T.~Akidau, A.~Balikov, K.~Bekiro{\u{g}}lu, S.~Chernyak, J.~Haberman, R.~Lax,
  S.~McVeety, D.~Mills, P.~Nordstrom, and S.~Whittle.
\newblock Millwheel: fault-tolerant stream processing at internet scale.
\newblock {\em Proceedings of the VLDB Endowment}, 6(11):1033--1044, 2013.

\bibitem{dataflow}
T.~Akidau, R.~Bradshaw, C.~Chambers, S.~Chernyak, R.~J. Fernández-Moctezuma,
  R.~Lax, S.~McVeety, D.~Mills, F.~Perry, E.~Schmidt, and S.~Whittle.
\newblock {The Dataflow Model: A Practical Approach to Balancing Correctness,
  Latency, and Cost in Massive-Scale, Unbounded, Out-of-Order Data Processing}.
\newblock {\em Proceedings of the VLDB Endowment}, 8:1792--1803, 2015.

\bibitem{Ananthanarayanan:2013:PFS:2463676.2465272}
R.~Ananthanarayanan, V.~Basker, S.~Das, A.~Gupta, H.~Jiang, T.~Qiu,
  A.~Reznichenko, D.~Ryabkov, M.~Singh, and S.~Venkataraman.
\newblock Photon: Fault-tolerant and scalable joining of continuous data
  streams.
\newblock In {\em Proceedings of the 2013 ACM SIGMOD International Conference
  on Management of Data}, SIGMOD '13, pages 577--588, New York, NY, USA, 2013.
  ACM.

\bibitem{Arasu:2004:LRS:1316689.1316732}
A.~Arasu, M.~Cherniack, E.~Galvez, D.~Maier, A.~S. Maskey, E.~Ryvkina,
  M.~Stonebraker, and R.~Tibbetts.
\newblock Linear road: A stream data management benchmark.
\newblock In {\em Proceedings of the Thirtieth International Conference on Very
  Large Data Bases - Volume 30}, VLDB '04, pages 480--491. VLDB Endowment,
  2004.

\bibitem{Carbone:2017:SMA:3137765.3137777}
P.~Carbone, S.~Ewen, G.~F\'{o}ra, S.~Haridi, S.~Richter, and K.~Tzoumas.
\newblock State management in apache flink\&reg;: Consistent stateful
  distributed stream processing.
\newblock {\em Proc. VLDB Endow.}, 10(12):1718--1729, Aug. 2017.

\bibitem{carbone2015lightweight}
P.~Carbone, G.~F{\'o}ra, S.~Ewen, S.~Haridi, and K.~Tzoumas.
\newblock Lightweight asynchronous snapshots for distributed dataflows.
\newblock {\em arXiv preprint arXiv:1506.08603}, 2015.

\bibitem{DBLP:journals/debu/CarboneKEMHT15}
P.~Carbone, A.~Katsifodimos, S.~Ewen, V.~Markl, S.~Haridi, and K.~Tzoumas.
\newblock Apache flink{\texttrademark}: Stream and batch processing in a single
  engine.
\newblock {\em {IEEE} Data Eng. Bull.}, 38(4):28--38, 2015.

\bibitem{jefferson1985virtual}
D.~R. Jefferson.
\newblock Virtual time.
\newblock {\em ACM Transactions on Programming Languages and Systems (TOPLAS)},
  7(3):404--425, 1985.

\bibitem{li2005semantics}
J.~Li, D.~Maier, K.~Tufte, V.~Papadimos, and P.~A. Tucker.
\newblock Semantics and evaluation techniques for window aggregates in data
  streams.
\newblock In {\em Proceedings of the 2005 ACM SIGMOD international conference
  on Management of data}, pages 311--322. ACM, 2005.

\bibitem{li2008out}
J.~Li, K.~Tufte, V.~Shkapenyuk, V.~Papadimos, T.~Johnson, and D.~Maier.
\newblock Out-of-order processing: a new architecture for high-performance
  stream systems.
\newblock {\em Proceedings of the VLDB Endowment}, 1(1):274--288, 2008.

\bibitem{Liu:2006:ROS:1142473.1142513}
B.~Liu, Y.~Zhu, and E.~Rundensteiner.
\newblock Run-time operator state spilling for memory intensive long-running
  queries.
\newblock In {\em Proceedings of the 2006 ACM SIGMOD International Conference
  on Management of Data}, SIGMOD '06, pages 347--358, New York, NY, USA, 2006.
  ACM.

\bibitem{10.1007/978-3-642-10424-4_16}
M.~R.~N. Mendes, P.~Bizarro, and P.~Marques.
\newblock A performance study of event processing systems.
\newblock In R.~Nambiar and M.~Poess, editors, {\em Performance Evaluation and
  Benchmarking}, pages 221--236, Berlin, Heidelberg, 2009. Springer Berlin
  Heidelberg.

\bibitem{murray2013naiad}
D.~G. Murray, F.~McSherry, R.~Isaacs, M.~Isard, P.~Barham, and M.~Abadi.
\newblock {Naiad: a timely dataflow system}.
\newblock In {\em Proceedings of the Twenty-Fourth ACM Symposium on Operating
  Systems Principles}, pages 439--455. ACM, 2013.

\bibitem{Noghabi:2017:SSS:3137765.3137770}
S.~A. Noghabi, K.~Paramasivam, Y.~Pan, N.~Ramesh, J.~Bringhurst, I.~Gupta, and
  R.~H. Campbell.
\newblock Samza: Stateful scalable stream processing at linkedin.
\newblock {\em Proc. VLDB Endow.}, 10(12):1634--1645, Aug. 2017.

\bibitem{DBLP:journals/corr/Ruder16}
S.~Ruder.
\newblock An overview of gradient descent optimization algorithms.
\newblock {\em CoRR}, abs/1609.04747, 2016.

\bibitem{Srivastava:2004:FTM:1055558.1055596}
U.~Srivastava and J.~Widom.
\newblock Flexible time management in data stream systems.
\newblock In {\em Proceedings of the Twenty-third ACM SIGMOD-SIGACT-SIGART
  Symposium on Principles of Database Systems}, PODS '04, pages 263--274, New
  York, NY, USA, 2004. ACM.

\bibitem{Stephens1997}
R.~Stephens.
\newblock A survey of stream processing.
\newblock {\em Acta Informatica}, 34(7):491--541, 1997.

\bibitem{to2017survey}
Q.-C. To, J.~Soto, and V.~Markl.
\newblock A survey of state management in big data processing systems.
\newblock {\em arXiv preprint arXiv:1702.01596}, 2017.

\bibitem{Tu:2006:LSS:1182635.1164195}
Y.-C. Tu, S.~Liu, S.~Prabhakar, and B.~Yao.
\newblock Load shedding in stream databases: A control-based approach.
\newblock In {\em Proceedings of the 32Nd International Conference on Very
  Large Data Bases}, VLDB '06, pages 787--798. VLDB Endowment, 2006.

\bibitem{YANG1996113}
Q.~Yang and H.~N. Koutsopoulos.
\newblock A microscopic traffic simulator for evaluation of dynamic traffic
  management systems.
\newblock {\em Transportation Research Part C: Emerging Technologies}, 4(3):113
  -- 129, 1996.

\end{thebibliography}
